\documentclass[pdflatex,sn-mathphys-ay]{sn-jnl}

\usepackage{graphicx}
\usepackage{multirow}
\usepackage{amsmath,amssymb,amsfonts}
\usepackage{amsthm}
\usepackage{mathrsfs}
\usepackage[title]{appendix}
\usepackage[dvipsnames]{xcolor}
\usepackage{textcomp}
\usepackage{manyfoot}
\usepackage{booktabs}
\usepackage{algorithm}
\usepackage{algorithmicx}
\usepackage{algpseudocode}
\usepackage{listings}
\usepackage[colorinlistoftodos]{todonotes}
\usepackage[dvipsnames]{xcolor}

\usetikzlibrary{calc}

\definecolor{yellow}{RGB}{252, 211, 3}
\definecolor{green}{RGB}{60,179,113}
\definecolor{blue}{RGB}{3, 165, 252}
\definecolor{gray}{RGB}{70, 80, 75}

\definecolor{lightblue}{rgb}{.90,.95,1}
\definecolor{lightgreen}{rgb}{.90,1,.95}
\definecolor{darkgreen}{rgb}{0,.5,0.5}
\definecolor{darkblue}{rgb}{0.122, 0.467, 0.706}
\definecolor{slightgray}{RGB}{160, 160, 160}

\theoremstyle{thmstyleone}

\theoremstyle{thmstyletwo}

\theoremstyle{thmstylethree}

\begin{document}

\title[Principled Data Assimilation of Turbulence]{Toward Principled Generative Data Assimilation of Turbulent Flows from Sparse Observations}

\author[1,2]{\fnm{Baris} \sur{Turan}}\email{baris.turan@itlr.uni-stuttgart.de}

\author[1,2]{\fnm{Zhuoran} \sur{Liu}}\email{zhuoran.liu@itlr.uni-stuttgart.de}

\author*[1,2]{\fnm{Heng} \sur{Xiao}}\email{heng.xiao@simtech.uni-stuttgart.de}

\affil*[1]{\orgdiv{Stuttgart Center for Simulation
Science}, \orgname{University of Stuttgart}, \city{Stuttgart}, \postcode{70569}, \country{Germany}}

\affil[2]{\orgdiv{Institute of Aerospace Thermodynamics}, \orgname{University of Stuttgart}, \city{Stuttgart}, \postcode{70569}, \country{Germany}}

\abstract{Turbulent flows are highly chaotic, which makes their instantaneous states difficult to predict. Data assimilation aims to reduce this predictive uncertainty by synthesizing the predictions of a physical solver with partial observations of the system. Traditional data assimilation methods such as ensemble and variational approaches can incur substantial computational cost through evaluations of high-fidelity solvers and, for variational methods, adjoint calculations. Recently, generative diffusion models have been explored in the solution of inverse problems in fluid mechanics, demonstrating their potential for data assimilation.
However, existing diffusion-based methods often use free parameters instead of the prescribed observation-error covariance to control the balance between the prior and likelihood, departing from the Bayesian formulation of data assimilation.
We propose a framework for principled generative data assimilation in which observations are assimilated in the clean-state space and the prescribed observation-error covariance enters explicitly through an ensemble Kalman update. In the Lorenz-63 system, the proposed method yields a lower trajectory error and equation residual than the other diffusion-based posterior sampling methods considered, although its ensemble underestimates the posterior uncertainty. In two-dimensional Rayleigh--B\'enard convection, the generated fields recover the statistics of the direct numerical simulation reference at large scales, with deviations confined to the smaller scales. These results suggest that the framework can produce physically plausible turbulent flow fields, indicating its potential for more challenging turbulent flow applications.
}

\maketitle
\section{Introduction}\label{sec:intro}
Turbulence is ubiquitous in natural and engineering flows, from atmospheric boundary layers and ocean circulation to combustion chambers and aerodynamic wakes. A central difficulty in predicting turbulent flows is their sensitive dependence on initial conditions: small perturbations are amplified through nonlinear interactions across scales, so that even a high-fidelity simulation has a finite predictability horizon, beyond which it no longer tracks the particular realization of the flow. Uncertain boundary and initial conditions shorten that horizon further. Such a limitation of the predictive horizon can be addressed by data assimilation (DA), in which two imperfect sources of information, the prediction of a physical solver and sparse observations of the system, are combined into a better estimate of the state than either provides alone. In Bayesian terms, the prediction of the solver is the prior, which the observations update to yield the posterior.

Conventional DA methods fall into three broad families: Monte Carlo, variational and ensemble-based methods~\mbox{\citep{bannister2017review}}. Monte Carlo methods such as the particle filter place no restriction on the prior or the observation operator, but the number of particles they require grows exponentially with the dimension of the problem, which limits their use in turbulent flows~\mbox{\citep{fearnhead2018particle}}. The other two families differ in how they obtain the prior covariance. Variational methods such as 3DVar and 4DVar find the mode of the posterior under Gaussian assumptions~\mbox{\citep{evensen2022data}}, usually with a background error covariance modeled from statistical estimates, and return a single estimate rather than a spread~\mbox{\citep{bannister2017review}}. Ensemble-based methods such as the ensemble Kalman filter (EnKF) instead estimate that covariance from an ensemble of states at every update step, so that it evolves with the flow~\mbox{\citep{evensen1994sequential}}. Operational systems combine the two in hybrid ensemble-variational schemes~\mbox{\citep{bannister2017review, whitaker2022comparison, bonavita2016evolution}}. Both families have been applied to turbulent flows. Variational methods have reconstructed channel-flow states from limited observations and inferred near-wall turbulence from outer-flow measurements~\mbox{\citep{wang2021state, wang2025variational}}. Ensemble methods have reconstructed near-wall states from wall measurements~\mbox{\citep{colburn2011state}}, inferred the Reynolds stresses of Reynolds-averaged Navier--Stokes (RANS) models from sparse velocity observations~\mbox{\citep{xiao2016quantifying}}, and assimilated experimental measurements into large-eddy simulations of turbulent jets~\mbox{\citep{labahn2020ensemble}}. For the present work, an important distinction between the two is how the prior covariance is represented.
Despite their different formulations, both are computationally demanding for turbulent flows because they require repeated evaluations of the physical solver. Ensemble methods require repeated solver evaluations across ensemble members, which is prohibitive when resolving the relevant scales demands many degrees of freedom~\mbox{\citep{vishny2024high, villanueva2025enhancement}}. Variational methods avoid large ensembles but require repeated forward and adjoint evaluations, and adjoint solvers are demanding to develop and maintain and are unavailable for many codes, such as proprietary or legacy solvers~\mbox{\citep{carrassi2018data, maulik2022efficient}}.

Machine learning provides two main routes to enhance data assimilation. The first replaces or corrects the physical solver with a learned model~\mbox{\citep{brajard2021combining, ozalp2026real}}. The second learns components of the assimilation procedure itself. Some approaches learn the analysis mapping or the representation in which the analysis is performed~\mbox{\citep{zhang2024novel, zhou2024bi, zhou2026neural, bach2026learning}}, whereas others learn quantities entering the update, such as the background error covariance~\mbox{\citep{lu2025unet}}. Some methods span both directions, with 4DVarNet as one example~\mbox{\citep{fablet2021learning}}. A learned surrogate introduces an additional source of error, and it may lose the numerical stability of the solver it replaces~\mbox{\citep{ozalp2026real}}. A further limitation is that deterministic surrogates, such as neural operators~\mbox{\citep{li2020fourier, li2024learning}}, can efficiently propagate individual states but do not by themselves represent a distribution over possible states. This distinction matters for data assimilation, where the objective is to infer a posterior distribution rather than a single state. Generative models provide such a distribution directly and are therefore particularly attractive as priors for data assimilation~\mbox{\citep{zhou2026ensemblegenerativefilteringsequential}}.

Among generative models, diffusion models provide a natural framework for incorporating observations into a generative process. They use a forward process that gradually corrupts training samples with noise and learn a reverse process that progressively removes this noise to generate new samples~\mbox{\citep{ho2020denoising, song2021scorebased}}. Observations can then be incorporated into the reverse process to condition the generated samples on the available data. 
Most approaches impose this conditioning through guidance, in which a pretrained model is steered during generation towards samples consistent with the observations~\mbox{\citep{chung2025diffusionmodelsinverseproblems}}, without the task-specific retraining required by classifier-free guidance~\mbox{\citep{ho2021classifierfree}}. The most established scheme is diffusion posterior sampling (DPS;~\mbox{\citealp{chung2023diffusion}}). The likelihood of the observations given a noisy intermediate state is intractable, so DPS evaluates it at a single point estimate of the clean field obtained from the denoising network. \mbox{\citet{steinbrenner2026turbulence}} applied DPS to a latent diffusion model of turbulent plane Couette flow and observed two failure modes. With very sparse observations, the conditioning strength, which is normalized by the small observation residual, overweighted isolated observations and produced unphysical fields. With observations concentrated in a small region, the learned prior was distorted and the higher-order statistics degraded, although the samples matched the data locally. These failures expose a limitation of the point-estimate approximation: it discards the uncertainty of the clean field, particularly at high noise levels. In practical DPS implementations, the likelihood guidance is further rescaled for numerical stability, so that the relative influence of the observations is controlled by a tunable conditioning strength rather than directly by the prescribed observation-error covariance.

Several approaches have been developed to overcome the limitations of diffusion posterior sampling. One relaxes the point-estimate approximation. Pseudoinverse-guided diffusion models ($\Pi$GDM;~\mbox{\citealp{song2023pseudoinverse}}) replace the point estimate by an isotropic Gaussian whose covariance depends on the pseudo-time. Score-based data assimilation (SDA;~\mbox{\citealp{rozet2023score}}) uses a Gaussian of the same kind, whose covariance scales with the noise level and is set to a multiple of the identity in practice, and has been applied to weather and ocean state estimation~\mbox{\citep{qu2024deep, manshausen2025generative, martin2025generative}}. The other combines diffusion guidance with established data assimilation or inference techniques. Sequential unbiased resampling via Girsanov estimation (SURGE;~\mbox{\citealp{wei2026surge}}) incorporates a particle-filter update, while data assimilation with inverse sampling using stochastic interpolants (DAISI;~\mbox{\citealp{andrae2026daisi}}) pairs learned guidance with a dynamic forecast model. Ensemble Kalman guidance (EnKG;~\mbox{\citealp{zheng2025ensemble}}) requires only black-box access to the forward model. In these formulations, the weight given to the observations is set by a tunable parameter or by an assumed prior covariance, rather than by the prescribed observation-error covariance. We refer to a scheme in which this weighting follows from the prescribed observation-error covariance, as in the Bayesian formulation of DA, as \emph{principled data assimilation}. This is the property that the present work aims to recover.

We therefore build on decoupled annealed posterior sampling (DAPS;~\mbox{\citealp{zhang2025improving}}), which performs assimilation in the clean-state space, where the observation likelihood is naturally defined. DAPS in its original form assumes an isotropic Gaussian prior at each annealing step. We find that this yields posteriors that are too broad when observations are sparse, and we therefore estimate the prior covariance from an ensemble instead of prescribing it. We call the resulting method DAPS with an ensemble-based prior covariance (DAPS-e). Unlike the DPS-based approach of~\mbox{\citet{steinbrenner2026turbulence}}, DAPS-e applies the guidance to the clean state rather than to the noisy state. The prescribed observation-error covariance enters explicitly through the Kalman update and thus sets the balance between prior and likelihood, whereas practical DPS implementations control this balance through an additional tunable parameter~\mbox{\citep{chung2023diffusion}}. The method is tested on two cases with distinct roles. The Lorenz-63 system is the benchmark for posterior sampling: there the proposed method yields more accurate and dynamically consistent posterior samples than DPS, although its ensemble underestimates the posterior uncertainty. Rayleigh--B\'enard convection demonstrates the method on a turbulent flow, where the posterior samples reproduce key direct numerical simulation (DNS) statistics and remain physically plausible after assimilation. The paper is organized as follows. Section~\mbox{\ref{sec:methodology}} presents the method, Section~\mbox{\ref{sec:setup}} the two test cases, and Section~\mbox{\ref{sec:results}} the results; Section~\mbox{\ref{sec:conclusion}} concludes.

\section{Methodology}\label{sec:methodology}

In a conventional ensemble-based DA method, the state estimate is corrected in cycles. An ensemble of states is propagated forward by solving the governing equations over a fixed interval, each member starting from a different realization of the uncertain initial condition, so that the spread of the ensemble represents the uncertainty in the forecast. These forecast states form the prior. Measurements~$\boldsymbol{\Psi}$ taken over that interval, for example temperatures recorded at a set of probe locations, are then combined with the prior to give the posterior, and the cycle repeats. Almost all of the cost lies in propagating the ensemble, since every member requires a full solution of the governing equations.

In the present work, that propagation is replaced by a generative model. A collection of trajectories~$\boldsymbol{z}^{1:L}=\{\boldsymbol{z}^1,\dots,\boldsymbol{z}^L\}$, each a sequence of~$L$ states, is obtained once and offline from a solver, and a diffusion model is trained on it to learn the distribution from which the trajectories were drawn. At inference, the model is asked for a trajectory that is both a plausible evolution of the flow, which the learned distribution enforces, and consistent with the measurements~$\boldsymbol{\Psi}$ in the same Bayesian sense as in the conventional method. No governing equation is solved at this stage. Since the observations along the whole trajectory are assimilated together, the problem is one of smoothing rather than filtering.

\begin{figure}
    \centering
    \includegraphics[width=\linewidth]{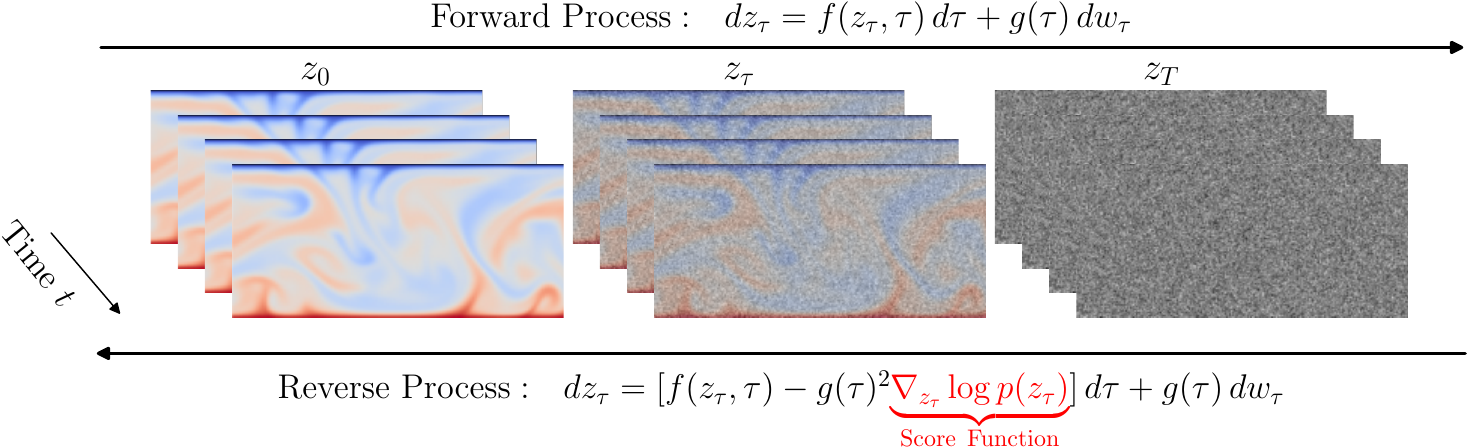}
    \caption{Schematic of the forward and reverse processes of a diffusion model, shown for the Rayleigh--B\'enard convection case. In the forward process, the clean sequence of flow fields~$\boldsymbol{z}_0$ is corrupted with noise at each pseudo-time~$\tau$, giving the noisy fields~$\boldsymbol{z}_\tau$. At~$\tau=T$, the noisy field~$\boldsymbol{z}_T$ is indistinguishable from noise.}
    \label{fig:diffusion_schematic}
\end{figure}

Taking the Rayleigh--B\'enard case of Section~\mbox{\ref{sec:setup}} as the example,~$\boldsymbol{z}_0$ is a short sequence of temperature fields (Figure~\mbox{\ref{fig:diffusion_schematic}}). A diffusion model corrupts such a field with Gaussian noise in many small steps until only noise remains, and learns to undo that corruption. The subscript indicates how far along this corruption a field lies:~$\boldsymbol{z}_\tau$ is part signal and part noise, and~$\boldsymbol{z}_T$ is noise alone. Note that the pseudo-time~$\tau$ is distinct from the physical time~$t$ along which the trajectory evolves. Section~\mbox{\ref{subsec:diffusion_models}} describes how a plausible~$\boldsymbol{z}_0$ is generated from~$\boldsymbol{z}_T$, and Section~\mbox{\ref{subsec:da_diffusion}} how it is made to agree with the measurements. The state is written generally as~$\boldsymbol{z}$, since the same formulation is applied to the Lorenz-63 system, where~$\boldsymbol{z}$ is a three-component vector.

\subsection{Diffusion Models}\label{subsec:diffusion_models}
Diffusion models
consist of a prescribed forward process, which incrementally adds noise to a clean sample~$\boldsymbol{z}_0$ from the training data, and a learned reverse process, which removes it.
The forward process is described by the stochastic differential equation~(SDE):
\begin{equation}
    d\boldsymbol{z}_\tau
    =
    \boldsymbol{f}(\boldsymbol{z}_\tau,\tau)d\tau
    +
    \boldsymbol{g}(\tau)d\boldsymbol{w}_\tau,
    \quad \tau\in[0,T],
\end{equation}
where~$\tau$ is the diffusion pseudo-time, which parameterizes the noise level
from the clean state at~$\tau=0$ to the noisiest state at~$\tau=T$;
$\boldsymbol{f}$ and~$\boldsymbol{g}$ are the drift and diffusion coefficients,
and~$\boldsymbol{w}_\tau$ is Brownian motion.
Its transition kernel is Gaussian,
\begin{equation}\label{eq:forward_kernel}
    p(\boldsymbol{z}_\tau|\boldsymbol{z}_0)=\mathcal{N}\left(\boldsymbol{z}_\tau |\mu_\tau\boldsymbol{z}_0, \sigma_\tau^2\boldsymbol{I}\right),
\end{equation}
where~$\mu_\tau$ and~$\sigma_\tau$ follow from~$\boldsymbol{f}$ and~$\boldsymbol{g}$ and are chosen such that~$p(\boldsymbol{z}_T)\approx\mathcal{N}(0, \sigma_T^2\boldsymbol{I})$, so that no trace of the original field survives at~$\tau=T$.
Generation runs this process backwards along the reverse SDE:
\begin{equation}\label{eq:reverse_sde}
    d\boldsymbol{z}_\tau
    =
    \left[
        \boldsymbol{f}(\boldsymbol{z}_\tau,\tau)
        -
        \boldsymbol{g}(\tau)^2
        \nabla_{\boldsymbol{z}_\tau}
        \log p(\boldsymbol{z}_\tau)
    \right]d\tau
    +
    \boldsymbol{g}(\tau)d\boldsymbol{w}_\tau,
\end{equation}
which requires the score~$\nabla_{\boldsymbol{z}_\tau}\log p(\boldsymbol{z}_\tau)$, the gradient of the log-density at pseudo-time~$\tau$. The score is unknown and is approximated by a neural network~$\boldsymbol{s}_\phi(\boldsymbol{z}_\tau,\tau)$. The network is trained on pairs of clean and noisy states produced by the forward process: the forward process supplies the training data, and the reverse process uses what was learned. We use the noise-prediction objective of the denoising diffusion probabilistic model (DDPM; \citealp{ho2020denoising}), given in Appendix~\mbox{\ref{sec:A1}}.

\subsection{Data Assimilation with Diffusion Models}\label{subsec:da_diffusion}
We seek the posterior of the trajectory,~$p(\boldsymbol{z}\mid\boldsymbol{\Psi})\propto p(\boldsymbol{\Psi}\mid\boldsymbol{z})\,p(\boldsymbol{z})$, where the superscript~$1:L$ is dropped from here on and the dependence of~$\boldsymbol{z}$ on space and physical time is implied. The observations are related to the state by
\begin{equation}
    \boldsymbol{\Psi}=\mathcal{H}(\boldsymbol{z})+\boldsymbol{\varepsilon}
\end{equation}
where~$\mathcal{H}$ is the observation operator and~$\boldsymbol{\varepsilon}\sim\mathcal{N}(0, \boldsymbol{R})$ the observation noise, with covariance~$\boldsymbol{R}=\mathrm{diag}(\sigma_{\mathrm{obs},i}^2)$. Sampling the posterior with a diffusion model requires the posterior score~$\nabla_{\boldsymbol{z}_\tau} \log p(\boldsymbol{z}_\tau\mid\boldsymbol{\Psi})$. A network can be trained to approximate it directly from paired data~$(\boldsymbol{z}, \boldsymbol{\Psi})$~\mbox{\citep{shysheya2024conditional}}, but a separate network is then needed for every observation setup. Training-free guidance schemes instead use Bayes' theorem to decompose the posterior score,
\begin{equation}
\nabla_{\boldsymbol{z}_\tau}\log{p(\boldsymbol{z}_\tau\mid\boldsymbol{\Psi})}=\nabla_{\boldsymbol{z}_\tau}\log{p(\boldsymbol{z}_\tau)} +\nabla_{\boldsymbol{z}_\tau} {\log p(\boldsymbol{\Psi}\mid\boldsymbol{z}_\tau)}.
\label{eq:dps_bayes}
\end{equation}
The first term is the prior score learned by the diffusion model. The second, the likelihood score, is intractable except at~$\tau=0$, because the observations constrain the clean state and not the noisy one:
\begin{equation}
    p(\boldsymbol{\Psi}|\boldsymbol{z}_\tau)=\int p(\boldsymbol{\Psi} \vert \boldsymbol{z}_0,\boldsymbol{z}_\tau)p(\boldsymbol{z}_0 \vert \boldsymbol{z}_\tau)\,d\boldsymbol{z}_0=\int p(\boldsymbol{\Psi} \vert \boldsymbol{z}_0)p(\boldsymbol{z}_0 \vert \boldsymbol{z}_\tau)\,d\boldsymbol{z}_0.
\end{equation}
The likelihood~$p(\boldsymbol{\Psi}\mid\boldsymbol{z}_0)$ follows from the observation model, but~$p(\boldsymbol{z}_0\mid\boldsymbol{z}_\tau)$ has no closed form. In the terms of Figure~\mbox{\ref{fig:diffusion_schematic}}, it is the set of clean temperature fields that could have produced the partly corrupted field~$\boldsymbol{z}_\tau$, each weighted by how plausible it is. It is wide when~$\boldsymbol{z}_\tau$ is close to noise and narrow when~$\boldsymbol{z}_\tau$ is nearly clean. How this conditional is approximated is what distinguishes the two schemes below.

\subsubsection{Diffusion Posterior Sampling}\label{subsec:da_with_diffusion_models}
Diffusion posterior sampling (DPS;~\mbox{\citealp{chung2023diffusion}}) replaces~$p(\boldsymbol{z}_0\mid\boldsymbol{z}_\tau)$ by a point mass at the denoised estimate~$\hat{\boldsymbol{z}}_0=\mathbb{E}[\boldsymbol{z}_0\mid\boldsymbol{z}_\tau]$, which is given by Tweedie's formula~\mbox{\citep{efron2011tweedie}}
\begin{equation}\label{eq:tweedie}
    \hat{\boldsymbol{z}}_0
    =
    \frac{
        \boldsymbol{z}_\tau
        +
        \sigma_\tau^2
        \nabla_{\boldsymbol{z}_\tau}
        \log p(\boldsymbol{z}_\tau)
    }{
        \mu_\tau
    }
    \approx
    \frac{
        \boldsymbol{z}_\tau
        +
        \sigma_\tau^2
        \boldsymbol{s}_\phi(\boldsymbol{z}_\tau,\tau)
    }{
        \mu_\tau
    }.
\end{equation}
Physically, DPS treats the single best estimate of the clean field as the only field that could have produced~$\boldsymbol{z}_\tau$. This is harmless when~$\boldsymbol{z}_\tau$ is nearly clean, and least justified at high noise levels, where many different clean fields are consistent with~$\boldsymbol{z}_\tau$ and the guidance is most likely to mislead. For isotropic observation noise,~$\boldsymbol{R}=\sigma_{\mathrm{obs}}^2\boldsymbol{I}$, the likelihood score then becomes
\begin{equation}
\label{eq:dps_first}
    \nabla_{\boldsymbol{z}_\tau} {\log p(\boldsymbol{\Psi}|\boldsymbol{z}_\tau)}\approx-\frac{1}{2\sigma_{\mathrm{obs}}^2}\nabla_{\boldsymbol{z}_\tau}\lVert\boldsymbol{\Psi}-\mathcal{H}(\boldsymbol{\hat{z}}_0(\boldsymbol{z}_\tau))\rVert^2.
\end{equation}
In practice, directly using $1/(2\sigma_{\mathrm{obs}}^2)$ to scale the likelihood gradient can lead to unstable guidance. Following~\citet{chung2023diffusion}, we instead use a conditioning strength normalized by the observation residual,
\begin{equation}\label{eq:cond_strength}
\nabla_{\boldsymbol{z}_\tau}
\log p(\boldsymbol{\Psi}\mid\boldsymbol{z}_\tau)
\approx
-\xi
\nabla_{\boldsymbol{z}_\tau}
\left\|
\boldsymbol{\Psi}
-
\mathcal{H}(\hat{\boldsymbol{z}}_0)
\right\|^2,
\qquad
\xi
=
\frac{\zeta}{
\left\|
\boldsymbol{\Psi}
-
\mathcal{H}(\hat{\boldsymbol{z}}_0)
\right\|
}
\end{equation}
where $\zeta$ is a tunable hyperparameter.
The reason for this normalization can be understood by examining the gradient of the squared residual in Equation~\ref{eq:dps_first}. This gradient is proportional to the residual itself. Early in the reverse process~$\hat{\boldsymbol{z}}_0$ is still far from any data-consistent state, the residual is large, and the update is large enough to destabilize the generation. Near the end the residual is small and the update would vanish before the observations had been enforced. Dividing by the residual norm holds the guidance at a roughly constant strength.
The difficulty is that the normalization responds to the size of the residual and not to its cause. When observations are sparse, the residual is small because it is assembled from few terms, even if~$\hat{\boldsymbol{z}}_0$ is far from the true state, and dividing by it overweights the few observations together with their noise. This is the first failure mode reported by~\mbox{\citet{steinbrenner2026turbulence}} (Section~\mbox{\ref{sec:intro}}).
In conventional DA, the observation-error covariance~$\boldsymbol{R}$ would set the balance between prior and observations. In the DPS formulation used here, that balance is set by a stability requirement through~$\xi$ instead. Variants of DPS address this in different ways; the point made here concerns this formulation only.

\subsubsection{Decoupled Annealed Posterior Sampling}
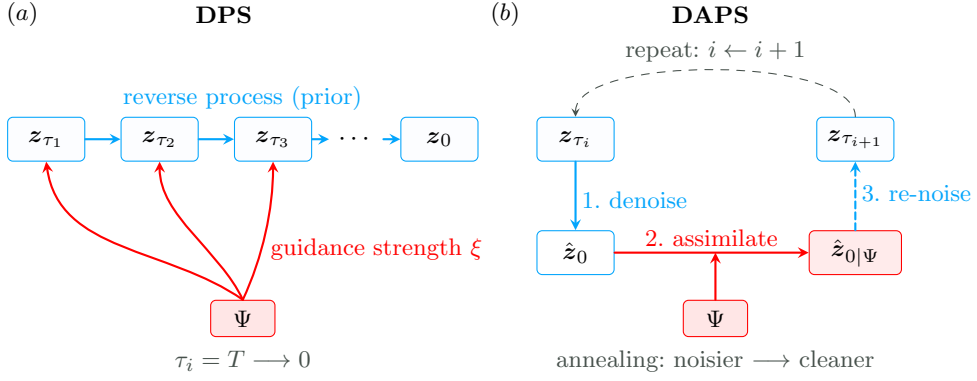
\begin{figure}
\centering
\begin{tikzpicture}[
x=1cm, y=1cm,
>=stealth,
line cap=round,
line join=round,
prior/.style={blue, thick},
data/.style={red, thick},
state/.style={
draw=blue, fill=lightblue!12,
rounded corners=2pt,
minimum width=1.02cm, minimum height=0.58cm,
inner sep=1.6pt, align=center, font=\small
},
clean/.style={
draw=blue, fill=white,
rounded corners=2pt,
minimum width=1.02cm, minimum height=0.58cm,
inner sep=1.6pt, align=center, font=\small
},
analysis/.style={
draw=red, fill=red!8,
rounded corners=2pt,
minimum width=1.22cm, minimum height=0.58cm,
inner sep=1.6pt, align=center, font=\small
},
obs/.style={
draw=red, fill=red!10,
rounded corners=2pt,
minimum width=0.85cm, minimum height=0.48cm,
inner sep=1.4pt, align=center, font=\small
},
steplab/.style={
font=\scriptsize, text=blue, inner sep=1pt, fill=white
},
datalab/.style={
font=\scriptsize, text=red, inner sep=1pt, fill=white
},
iterlab/.style={
font=\scriptsize, text=gray, inner sep=1pt, fill=white
}
]


\node[anchor=west, font=\small\bfseries] at (-0.50,3.6) {$(a)$};
\node[anchor=west, font=\small\bfseries] at (2.0,3.6) {DPS};

\node[state] (dT) at (0.15,1.95) {$\boldsymbol{z}_{\tau_1}$};
\node[state] (dt) at (1.65,1.95) {$\boldsymbol{z}_{\tau_2}$};
\node[state] (dtm) at (3.15,1.95) {$\boldsymbol{z}_{\tau_3}$};
\node (ddots) at (4.25,1.95) {$\cdots$};
\node[state] (d0) at (5.35,1.95) {$\boldsymbol{z}_0$};

\draw[->, prior] (dT) -- (dt);
\draw[->, prior] (dt) -- (dtm);
\draw[->, prior] (dtm) -- (ddots);
\draw[->, prior] (ddots) -- (d0);

\node[font=\scriptsize, text=blue] at (2.75,2.5)
    {reverse process (prior)};

\node[obs] (dy) at (2.75,-0.42) {$\Psi$};

\draw[->, data] (dy.north)
    to[out=145, in=-82, looseness=0.98] (dT.south);
\draw[->, data] (dy.north)
    to[out=115, in=-90, looseness=0.86] (dt.south);
\draw[->, data] (dy.north)
    to[out=65, in=-90, looseness=0.86] (dtm.south);

\node[datalab] at (4.52,0.5)
    {guidance strength~$\xi$};

\node[font=\scriptsize, text=gray] at (2.75,-0.98)
    {$\tau_i=T\longrightarrow 0$};


\node[anchor=west, font=\small\bfseries] at (5.9,3.6) {$(b)$};
\node[anchor=west, font=\small\bfseries] at (8.3,3.6) {DAPS};

\node[state] (zti) at (7.15,1.95)
    {$\boldsymbol{z}_{\tau_i}$};

\node[state] (ztim1) at (10.85,1.95)
    {$\boldsymbol{z}_{\tau_{i+1}}$};

\node[clean] (zhat) at (7.15,0.45)
    {$\hat{\boldsymbol{z}}_0$};

\node[analysis] (zPsi) at (10.85,0.45)
    {$\hat{\boldsymbol{z}}_{0\mid\Psi}$};

\node[obs] (aPsi) at (9.00,-0.42)
    {$\Psi$};

\draw[->, prior] (zti) -- (zhat)
    node[steplab, pos=0.55, right=1pt]
    {1. denoise};

\draw[->, data] (zhat) -- (zPsi)
    node[datalab, pos=0.5, above=1pt]
    {2. assimilate};

\draw[->, prior, dash pattern=on 2.6pt off 1.5pt]
    (zPsi) -- (ztim1)
    node[steplab, pos=0.55, right=1pt]
    {3. re-noise};

\draw[->, data]
    (aPsi) -- ($(zhat)!0.5!(zPsi)$);

\draw[->, gray, dashed]
    (ztim1.north)
    .. controls +(0,0.82) and +(0,0.82) ..
    (zti.north)
    node[midway, above=2pt, iterlab]
    {repeat:~$i\leftarrow i+1$};

\node[font=\scriptsize, text=gray, align=center]
    at (9.00,-0.98)
    {annealing: noisier~$\longrightarrow$ cleaner};

\end{tikzpicture}

\caption{
An illustration of~$(a)$ diffusion posterior sampling (DPS) and~$(b)$
decoupled annealed posterior sampling (DAPS). In DPS, guidance is added to
the noisy field~$\boldsymbol{z}_{\tau_i}$, with a tunable hyperparameter
controlling the guidance strength. In DAPS, the assimilation is performed on
the estimated clean state~$\hat{\boldsymbol{z}}_0$ instead of the noisy state.
The dashed gray arrow indicates advancement to the next annealing iteration:
$\boldsymbol{z}_{\tau_{i+1}}$ is used as the current state in the next step,
without an additional transformation.
}
\label{fig:methodology}
\end{figure}

Decoupled annealed posterior sampling (DAPS;~\mbox{\citealp{zhang2025improving}})
moves the assimilation to the clean state, where the likelihood is defined, so
that the prescribed observation-error covariance can enter the update directly
(Figure~\mbox{\ref{fig:methodology}}).
For diffusion pseudo-times~$\tau_1=T>\tau_2>\dots>0$, we define
$\sigma_{\mathrm{eff},\tau}=\sigma_\tau/\mu_\tau$ such that~$\boldsymbol{z}_\tau/\mu_\tau
=\boldsymbol{z}_0+\sigma_{\mathrm{eff},\tau}\boldsymbol{\epsilon}$, where~$\boldsymbol{\epsilon}\sim\mathcal{N}(\boldsymbol{0},\boldsymbol{I})$,
with~$\sigma_{\mathrm{eff},\tau_1}>
\sigma_{\mathrm{eff},\tau_2}>\dots>0$.
At each annealing step, DAPS first denoises the current state
$\boldsymbol{z}_{\tau_i}$ to a clean estimate~$\hat{\boldsymbol{z}}_0$ by
integrating the probability flow ordinary differential equation (ODE) of the
reverse process in~$n_{\mathrm{ODE}}$ steps, where~$n_{\mathrm{ODE}}=1$
recovers the Tweedie estimate of Equation~\mbox{\ref{eq:tweedie}}.
It then assimilates the observations by sampling
$\hat{\boldsymbol{z}}_{0\mid\boldsymbol{\Psi}}
\sim p(\boldsymbol{z}_0\mid\boldsymbol{z}_{\tau_i},\boldsymbol{\Psi})$
and re-noises the resulting sample to the next pseudo-time~$\tau_{i+1}$.
If~$\boldsymbol{z}_{\tau_i}$ follows
$p(\boldsymbol{z}_{\tau_i}\mid\boldsymbol{\Psi})$, this re-noising step yields
a sample from~$p(\boldsymbol{z}_{\tau_{i+1}}\mid\boldsymbol{\Psi})$.
Since~$p(\boldsymbol{z}_{\tau_1}\mid\boldsymbol{\Psi})
\approx p(\boldsymbol{z}_{\tau_1})$, progressively reducing the effective noise
level from~$\sigma_{\mathrm{eff},\tau_1}$ to zero therefore yields samples from
the posterior after~$N_{\mathrm{anneal}}$ annealing steps.

The assimilation step is defined as a Bayesian inference problem:
\begin{equation}
    p(\boldsymbol{z}_0\mid\boldsymbol{z}_\tau,\,\boldsymbol{\Psi})
    \propto
    p(\boldsymbol{z}_0\mid\boldsymbol{z}_\tau)
    p(\boldsymbol{\Psi}\mid\boldsymbol{z}_0).
    \label{eq:daps_bayes}
\end{equation}
with prior~$p(\boldsymbol{z}_0\mid\boldsymbol{z}_\tau)$ and
likelihood~$p(\boldsymbol{\Psi}\mid\boldsymbol{z}_0)$.
DAPS assumes an isotropic Gaussian prior
$p(\boldsymbol{z}_0\mid\boldsymbol{z}_\tau)
=\mathcal{N}(\hat{\boldsymbol{z}}_0,r_\tau^2\boldsymbol{I})$
with a tunable standard deviation~$r_\tau$. The resulting posterior is sampled
using Langevin dynamics, which iteratively combines the posterior gradient with
stochastic perturbations to generate conditional samples; further details are
given by~\citet{zhang2025improving}.

\subsubsection{DAPS with an Ensemble-Based Prior Covariance}\label{subsec:daps_e}
In the present experiments, the isotropic prior of DAPS produced posteriors that were too broad, particularly when observations were sparse. Its consequences depend on how strongly the data constrain the posterior. DAPS was developed and evaluated on image restoration tasks such as super-resolution, deblurring and inpainting~\mbox{\citep{zhang2025improving}}, in which every pixel, or a large fraction of them, enters the likelihood, and the likelihood then dominates the update. Here the observations cover a small fraction of the state, so the posterior is governed largely by the prior, whose assumed shape is no longer a detail. Moreover, an isotropic covariance carries no information about the spatial and temporal correlations of a turbulent field, which are precisely what should constrain the unobserved part of the state. We therefore replace the Langevin step by an ensemble Kalman update, in which the covariance of~$p(\boldsymbol{z}_0\mid\boldsymbol{z}_\tau)$ is estimated from an ensemble of clean-state estimates rather than prescribed. No other quantity passes from the ensemble into the sampler, and we refer to the scheme as DAPS with an ensemble-based prior covariance (DAPS-e).
Each member of the ensemble is updated as
\begin{equation}
    \hat{\boldsymbol{z}}_{0|\boldsymbol{\Psi}}^{(i)}
    =
    \hat{\boldsymbol{z}}_0^{(i)}
    +
    \boldsymbol{K}
    \left[
    \boldsymbol{\Psi}^{(i)}
    -
    \mathcal{H}\!\left(\hat{\boldsymbol{z}}_0^{(i)}\right)
    \right],
    \label{eq:enkf_update}
\end{equation}
where the Kalman gain~$\boldsymbol{K}$ is computed from the ensemble covariances (Appendix~\mbox{\ref{sec:A2}}). In the stochastic EnKF formulation used here, each member receives an independent realization~$\boldsymbol{\Psi}^{(i)}$ of the observations drawn according to~$\boldsymbol{R}$, which recovers the correct analysis covariance in expectation~\mbox{\citep{burgers1998analysis}}.

For the Lorenz-63 system, the raw update can exceed the noise level remaining at late annealing steps and push the sample away from regions that the learned prior represents well. We therefore bound the root mean square magnitude of the update by~$c\,\sigma_{\mathrm{eff},\tau}$, with a dimensionless constant~$c$, and leave it unchanged otherwise (Appendix~\mbox{\ref{sec:limiter}}). The limiter is a practical stabilization device, not a component derived from the sampler.

The parameter~$c$ sets the threshold of the DAPS-e update limiter, which bounds the assimilation update relative to the effective diffusion noise scale. It therefore bears some similarity to the conditioning-strength parameter~$\xi$ used in DPS, since both can influence the magnitude of the observational correction.
However, the two parameters are not doing the same kind of work. In the DPS formulation used here, the tuning of~$\xi$ is constrained by the stability of the guidance, which makes it difficult to let the prescribed observation-error covariance determine the weight given to the observations. In DAPS that weighting is more straightforward to obtain, since the covariance enters the update directly. In DAPS-e, the prescribed observation-error covariance enters explicitly through the Kalman gain and therefore determines the raw assimilation update before the limiter is applied. The parameter~$c$ acts only as a dimensionless stability factor that bounds this update relative to the effective diffusion noise scale fixed by the noise schedule. In contrast, the conditioning strength in DPS directly rescales the likelihood guidance and is not determined by the prescribed observation-error covariance. Thus, although~$c$ introduces an additional tunable parameter when the limiter is active, it does not replace the role of the observation-error covariance in determining the relative weighting of the prior and observations.
\section{Case Setup}\label{sec:setup}
\subsection{Lorenz-63 System}\label{subsec:setup_lorenz}
The Lorenz-63 system \citep{lorenz1963deterministic} is a simplified model of atmospheric convection. In this system, the evolution of the state vector~$\boldsymbol{z}=[z_1, z_2, z_3]^\top$ is governed by the following system of nonlinear ordinary differential equations:
\begin{align}
\frac{dz_1}{dt} &= \sigma_{\mathrm{L}} (z_2-z_1),\label{eq:lorenz1}\\
\frac{dz_2}{dt} &= z_1(\rho_{\mathrm{L}}-z_3)-z_2,\label{eq:lorenz2}\\
\frac{dz_3}{dt} &= z_1z_2-\beta_{\mathrm{L}} z_3.\label{eq:lorenz3}
\end{align}
For~$\sigma_{\mathrm{L}}=10,\,\rho_{\mathrm{L}}=28,\,\beta_{\mathrm{L}}=8/3$, the system exhibits chaotic dynamics along the well-known Lorenz attractor.
\begin{figure}[!t]
    \centering
    \includegraphics[width=\linewidth]{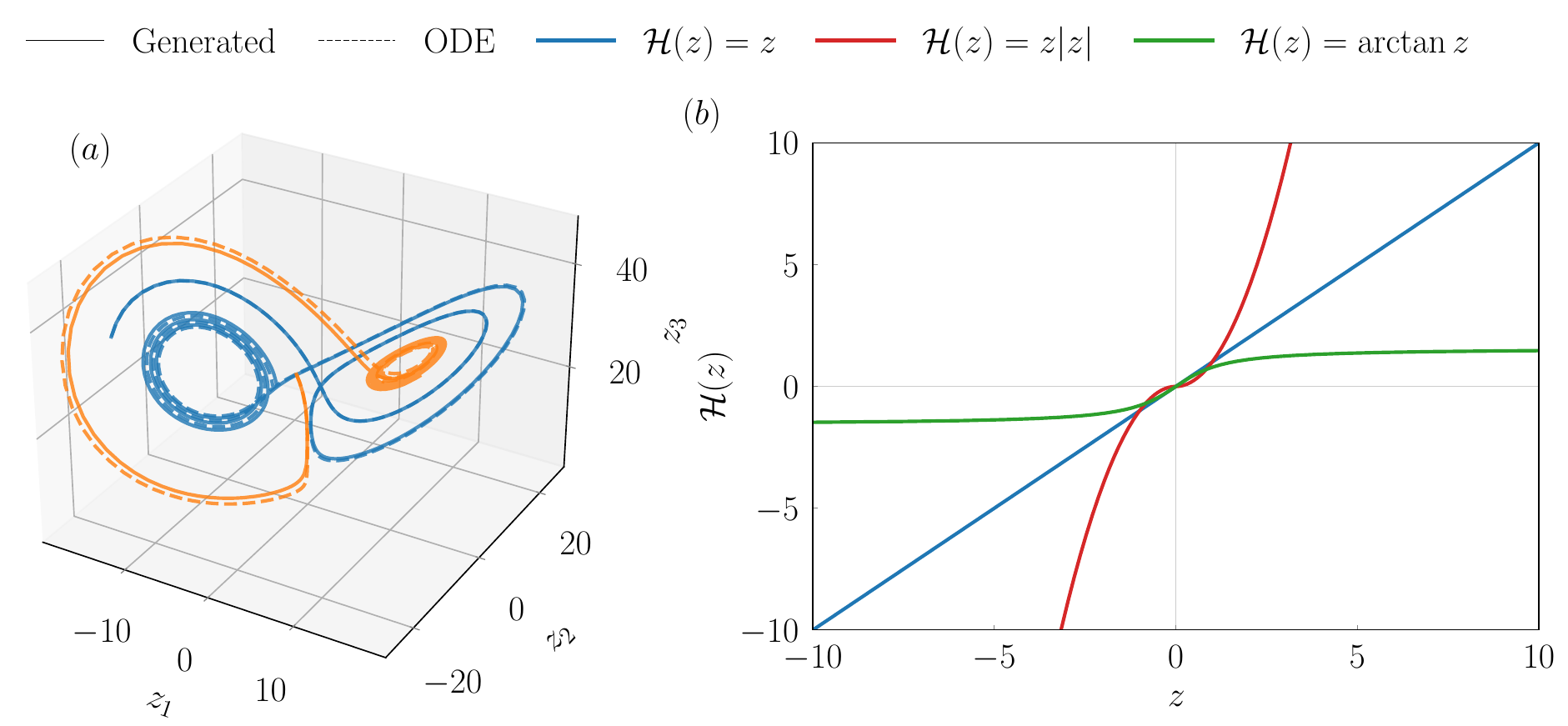}
    \caption{Setup for the Lorenz-63 system. Panel (a) shows two samples generated by the trained diffusion model together with trajectories obtained by integrating the Lorenz-63 equations from the same initial conditions using a fourth-order Runge--Kutta integrator. Panel (b) shows the three observation operators used for data assimilation. }
    \label{fig:trajectories-and-obs-operators}
\end{figure}
The training dataset for the Lorenz-63 system consists of 2000 trajectories, each initialized from a randomly sampled initial condition. The trajectories are obtained by integrating the governing equations using a fourth-order Runge--Kutta integrator with a timestep of~$\Delta t=0.01$ for 400 steps, after an initial transient of 1500 steps. We use a 1D UNet architecture for the diffusion model with residual blocks and sinusoidal pseudo-time embeddings. Additional details about the architecture and the training are provided in Appendix \ref{sec:A1}. Two trajectories generated by the trained model are shown in Figure \ref{fig:trajectories-and-obs-operators}, along with two trajectories obtained by integrating the Lorenz-63 equations from the same initial conditions as the generated ones. The generated trajectories show good agreement with those integrated from the same initial conditions and the butterfly-shaped Lorenz attractor is clearly visible in the figure.

Having established that the trained model reproduces the prior, we turn to the assimilation task itself, which is to recover a trajectory from sparse observations of it. We sample the observations from newly generated trajectories not present in the training dataset at a rate of 100$\Delta t=1.0$. This represents a substantially lower sampling frequency than those used in previous studies \citep{rozet2023score, andrae2026daisi, nagda2026generative}. We observe all three components of the state. We consider one linear and two nonlinear observation operators: the identity operator~$\mathcal{H}(\boldsymbol{z})=\boldsymbol{z}$, the signed square operator~$\mathcal{H}(\boldsymbol{z})=\boldsymbol{z}|\boldsymbol{z}|$ and the saturating arctangent operator~$\mathcal{H}(\boldsymbol{z})=\operatorname{arctan}(\boldsymbol{z})$. These operators are plotted in Figure~\ref{fig:trajectories-and-obs-operators}b. The signed square function arises in many quantities of interest in fluid mechanics. For instance, the
quadratic drag force on a body of cross-sectional area~$A$ and drag coefficient
$C_D$ is~$F_D = \tfrac{1}{2} \rho\, C_D A \, u \lvert u \rvert$, where~$\rho$ is
the density and~$u$ the velocity component along the direction of the force. The
signed form is retained because the drag opposes the direction of motion and must
reverse with it.
The arctangent operator, on the other hand, provides an idealized representation of sensor saturation, where the sensitivity of an instrument decays once the state leaves a finite band. In schlieren imaging, for example, the image contrast stops changing once the deflection of the light exceeds the measuring range set by the knife edge~\citep{saxena2025schlieren}.

Both the signed square and the arctangent operator pose difficulties for data assimilation. In both cases, the difficulty originates in the derivative of~$\mathcal{H}$, which
sets both how informative a measurement is and how strongly it steers the sampler. The derivative of the signed square function is~$\mathrm{d}\mathcal{H}/\mathrm{d}z = 2\lvert z
\rvert$, so the sensitivity of the operator grows without
bound with~$\lvert z \rvert$; over the Lorenz-63 attractor it spans several orders of
magnitude, whereas it is identically one for the identity operator. An error in the predicted state is therefore amplified before it enters the guidance, and a correction step
can overshoot even when the measurement itself is informative.
The derivative of the
arctangent operator,~$\mathrm{d}\mathcal{H}/\mathrm{d}z = 1/(1+z^2)$, is bounded by unity,
so it does not amplify the guidance in this way. However, it introduces a difficulty of a different kind, since the
sensitivity of the observation to changes in the state vanishes as~$\lvert z \rvert$
grows. This implies that the observations
only weakly constrain the state
beyond a finite band, and the posterior is dominated by the prior. This is a property
of the observing system rather than of any particular sampler. For gradient-guided schemes such as DPS there is a further consequence, in that the
likelihood term of Equation~\ref{eq:dps_first} vanishes together with the derivative,
so a conditioning strength~$\xi$ calibrated where the operator is responsive produces
correspondingly weak guidance where it saturates. The normalization of
Equation~\ref{eq:cond_strength} does not remedy this: it removes the dependence of the
update on the magnitude of the residual, but the sensitivity of the observation operator
survives it as a multiplicative factor in the guidance term. A single tuned~$\xi$
therefore cannot compensate for an operator whose sensitivity varies by orders of
magnitude.

We add heteroscedastic Gaussian noise to the observations, with the standard deviation of the noise as a function of the state~$\boldsymbol{z}(t)$ given by
\begin{equation}
    \sigma_{\mathrm{obs}} (\boldsymbol{z}(t))=0.1\lVert\mathcal{H}(\boldsymbol{z}(t))\rVert+0.05.
\end{equation}
In addition to DAPS-e, we run the original DAPS with Langevin dynamics and DPS for the Lorenz-63 case. In both DAPS and DAPS-e, we use~$N_{\text{anneal}}=200$ and~$n_{\text{ODE}}=5$, and in DAPS-e we set~$N_\mathrm{ens}=200$ and~$c=0.25$. In DAPS, we use 400 Langevin steps and set $r_\tau=\sigma_{\mathrm{eff}, \tau}$. We use~$\zeta=0.01$ in DPS. This~$\zeta$ value is tuned specifically for this problem and is an order of magnitude smaller than that recommended by~\citet{chung2023diffusion}.
For comparison, we also implement a bootstrap particle filter as a classical sequential Monte Carlo baseline, with~$1\times10^6$ particles for the identity operator and~$4\times10^6$ particles for the two nonlinear operators. We set the initial particle distribution to be a Gaussian centered around the ground truth. We add isotropic Gaussian jitter with a variance of $1\times10^{-4}$ to the particles during resampling. To obtain samples of the trajectory posterior, we trace each surviving particle back
through the resampling steps and assemble the states it passed through into a
complete trajectory. This
setup is susceptible to two related forms of degeneracy~\citep{fearnhead2018particle}.
First, at each observation time the importance weights tend to concentrate on a few
particles, so that only a small fraction of the ensemble contributes to the estimate. This degeneracy is known as weight degeneracy. Resampling counteracts it by
duplicating high-weight particles and discarding low-weight ones, but in doing so it
introduces path degeneracy: repeated resampling
progressively reduces the number of distinct ancestors at earlier times, so that the
reconstructed trajectories share a common history. We therefore do not treat the particle filter as a ground-truth or reference posterior, but rather as a conventional baseline for comparison with the diffusion-based samplers.

\subsection{Rayleigh--Bénard Convection}
\begin{figure}[!b]
    \centering
    \includegraphics[width=0.9\linewidth]{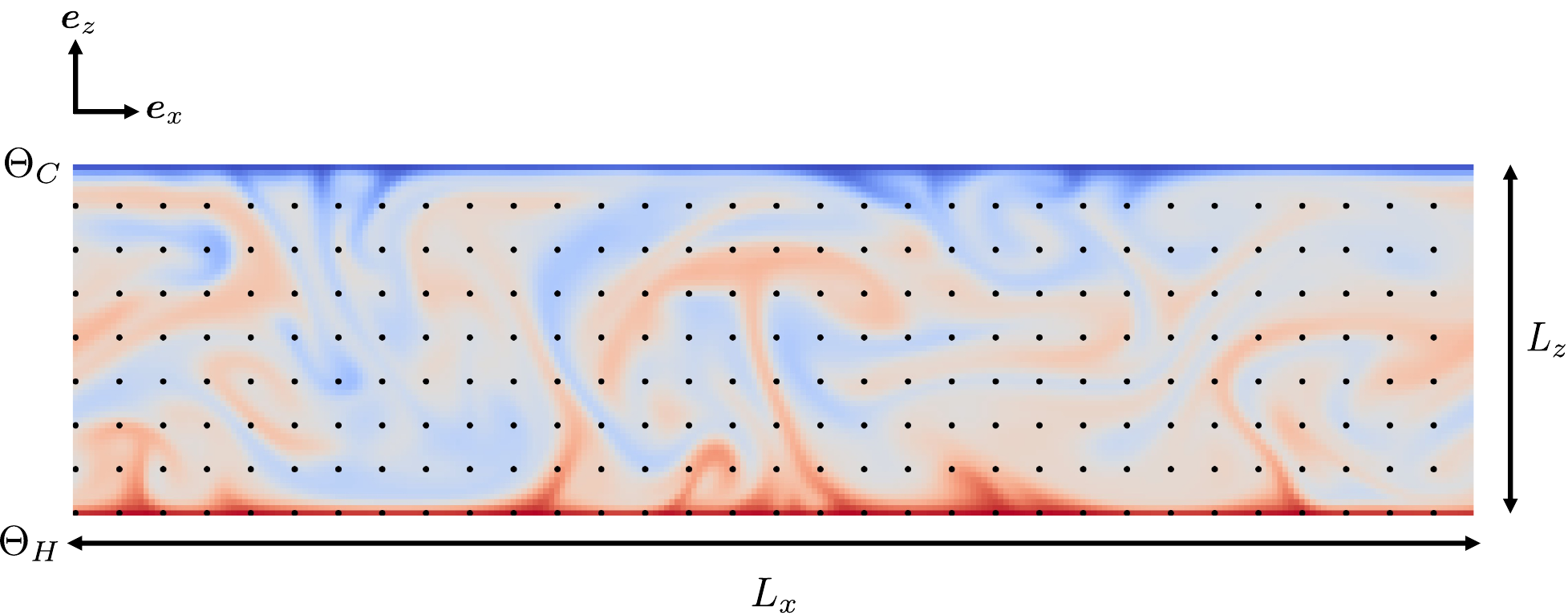}
    \caption{Temperature field from a Rayleigh--Bénard convection simulation. The bottom and top plates are held at constant temperatures~$\Theta_H$ and~$\Theta_C$, respectively, with~$\Theta_H>\Theta_C$. The observations are indicated with dots on the figure.}
    \label{fig:rbc_schematic}
\end{figure}

 Rayleigh--Bénard convection is a well-known example of natural convection encountered in many natural and engineering applications, such as convection in the atmosphere and the flow inside Earth's mantle~\citep{ahlers2009heat}. The Lorenz-63 system used as the first test case in this work is itself a severely truncated model of this problem, so the two cases considered here sit at opposite ends of the same reduction. As shown in Figure~\ref{fig:rbc_schematic}, it involves two parallel plates of width~$L_x$, separated by a distance of~$L_z$ in the vertical ($z$-) direction. The temperatures of the bottom and top plates are held constant at~$\Theta_H$ and~$\Theta_C$, respectively, with~$\Theta_H>\Theta_C$. The temperature gradient within the fluid between the two plates induces a density gradient, which in turn induces a buoyancy-driven flow. The hot fluid near the bottom plate has a  lower density and hence rises, while the cold, high-density fluid at the top sinks. This creates a cyclic convective flow pattern characterized by large plumes, as seen in Figure \ref{fig:rbc_schematic}.

Rayleigh--Bénard convection is usually modeled using the Boussinesq approximation, which neglects variation in density in the momentum equation for all terms except for the gravity term~$\rho g\,\boldsymbol{e}_z$. It assumes a linear relationship between density and temperature given by
\begin{equation}
    \rho(\Theta)=\rho_0-\alpha\rho_0(\Theta-\Theta_0),
\end{equation}
where~$\rho_0$ and~$\Theta_0$ are the reference density and temperature, respectively, and~$\alpha$ is the thermal expansion coefficient. The governing equations in non-dimensional form are given by
\begin{align}
\nabla\cdot\boldsymbol{u}&=0, \\
\frac{\partial \boldsymbol{u}}{\partial t}
+ (\boldsymbol{u}\cdot\nabla)\boldsymbol{u}
&=
-\nabla p
+ \theta\,\boldsymbol{e}_z
+ \sqrt{\frac{\mathrm{Pr}}{\mathrm{Ra}}}\,\nabla^2\boldsymbol{u},
\\
\frac{\partial \theta}{\partial t}
+ \boldsymbol{u}\cdot\nabla\theta
&=
\frac{1}{\sqrt{\mathrm{Ra} \cdot \mathrm{Pr}}}\,\nabla^2\theta,
\end{align}
where~$\boldsymbol{u}=[u_x, u_z]^\top$ is the velocity vector,~$p$ is the pressure, and~$\theta=\frac{\Theta-\Theta_C}{\Theta_H-\Theta_C}$ is the non-dimensional temperature. The Rayleigh number \(\mathrm{Ra}\) and the Prandtl number~$\mathrm{Pr}$ are the dimensionless parameters that characterize this problem. The Rayleigh number quantifies the relative strength of buoyancy-driven convection compared with the combined damping effects of viscous momentum diffusion and thermal diffusion, while the Prandtl number is the ratio of momentum diffusivity (kinematic viscosity) to thermal diffusivity.
At a critical Rayleigh number of order~$\mathrm{Ra}\sim10^3$, the conductive state becomes unstable and gives way to convection characterized by organized roll structures known as B\'enard cells. As the Rayleigh number increases further, the flow becomes increasingly unsteady and chaotic, eventually developing spatially disordered turbulent convection dominated by the emission and transport of thermal plumes~\citep{heslot1987transitions}.

The dataset consists of 1048 independent 2D DNS trajectories at~$\mathrm{Ra}=10^7$ and~$\mathrm{Pr}=1$, corresponding to the turbulent Rayleigh--B\'enard convection regime. We use 838 trajectories for training the model and reserve 105 trajectories for validation and another 105 for testing. The simulations are carried out using Dedalus~\citep{burns2020dedalus}, a flexible spectral open-source framework supporting various PDEs. Only the temperature field~$\theta$ is used for training. The computational domain has an aspect ratio of~$L_x/L_z=4$ and is discretized with a 512$\times$128 Cartesian grid along the~$x$- and~$z$-directions, respectively.
The simulation grid is stretched in the wall-normal direction to capture the sharp near-wall gradients. We interpolate the simulation outputs onto a uniform grid so that they are compatible with standard convolutional layers. We additionally coarsen the grid by a factor of 2 in both directions to decrease memory and computational overhead. The output is therefore generated on a 256$\times$64 grid. Periodic boundary conditions are used in the horizontal direction for both temperature and velocity. On the walls, no-slip boundary conditions are employed for velocity. The simulations are run for a total time of~$50t_f$ with a timestep of~$0.25t_f$, where~$t_f=\sqrt{\frac{L_z}{\alpha g(\Theta_H-\Theta_C)}}$ is the free fall time, or the time it takes for a fluid parcel to travel a distance~$L_z$ under the influence of buoyancy. We discard the initial transient of~$20t_f$ and train the model with the rest of the trajectories.
Each trajectory thus contains 120 snapshots and is initialized from a distinct initial temperature field consisting of a linear background perturbed by Gaussian noise:
\begin{equation}
    \theta(t=0) = \left(1-\frac{z}{L_z}\right) + \frac{z}{L_z}\left(1-\frac{z}{L_z} \right)\,\epsilon, \quad \epsilon\sim\mathcal{N}(0, 10^{-3}\boldsymbol{I}).
\end{equation}
The model is trained on short temporal sequences of snapshots rather than on the full 120-snapshot trajectory. During inference, the noise estimates obtained from these sequences are combined to form trajectories of arbitrary length, following the Markov blanket approach of~\citet{rozet2023score}, which is detailed in Appendix~\ref{sec:A1}. The trajectories generated in this way are 20 snapshots long.

The observations are sampled at every eighth grid point along each direction, so that 1.6\% of the grid points are observed. We use the identity operator~$\mathcal{H}(\boldsymbol{z})=\boldsymbol{z}$ and add Gaussian noise to the observations. The observations are sampled exclusively from the final~$5t_f$ of each trajectory, corresponding to the approximately statistically stationary portion of the flow evolution.  Each flow variable is decomposed into its average, denoted by an overline, and a fluctuation, denoted by a prime. We assimilate 16 independent trajectories from the test dataset in order to obtain sufficiently converged statistics. The posterior statistics
are averaged over the homogeneous horizontal direction, time, ensemble members,
and the independent test trajectories. The corresponding DNS statistics are
computed using the same spatial and temporal averaging procedure.
To ensure a fair comparison with the generated samples, all DNS statistics are
computed from the fields after interpolation onto the uniform grid and subsequent
coarsening to $256\times64$, rather than from the original $512\times128$
simulation fields.
We apply isotropic Gaussian noise to the observations with a standard deviation of $\sigma_\mathrm{obs} = 0.02$ in units of $\theta$, i.e., 2\% of the temperature difference between the plates. All 20 snapshots of the ground truth trajectory are observed. We use~$N_{\text{anneal}}=400$,~$n_\mathrm{ODE}=6$,~$N_\mathrm{ens}=80$, and~$c=0.25$ in DAPS-e.
The neural network architecture for this case is largely based on the UNet of  ~\citet{rozet2023score}. Additional details and hyperparameters are provided in Appendix~\ref{sec:A1}.

\section{Results}\label{sec:results}
\subsection{Lorenz-63 System}
\begin{figure}[!t]
    \centering
    \includegraphics[width=\linewidth]{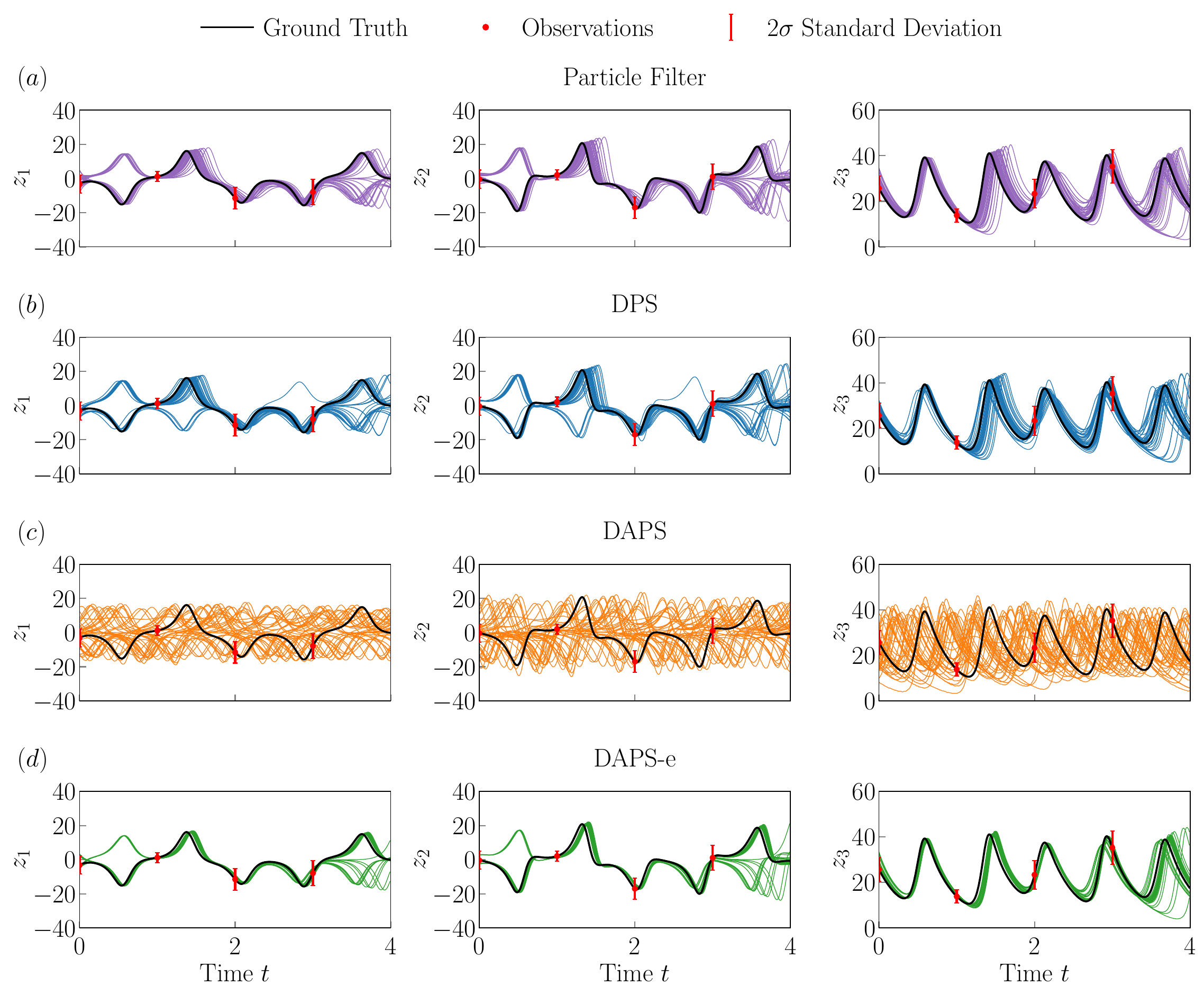}
    \caption{Representative posterior trajectories for the identity operator~$\mathcal{H}(z)=z$ generated by~$(a)$ the particle filter,~$(b)$ diffusion posterior sampling (DPS),~$(c)$ decoupled annealed posterior sampling (DAPS) and~$(d)$ the proposed method, DAPS with an ensemble-based prior covariance (DAPS-e). The error bars around the observations represent two standard deviations.}
    \label{fig:posterior_samples_linear}
\end{figure}
The posterior samples generated by DAPS-e show good overall agreement with the ground truth trajectory for the identity observation operator. We show the posterior trajectories generated by DAPS-e, DAPS and DPS in Figure~\ref{fig:posterior_samples_linear} for the identity operator, together with samples from the particle-filter baseline. DAPS-e captures the main overall trends in the posterior distribution, such as the bimodal nature of the distributions of~$z_1$ and~$z_2$. We can also see that the posterior distribution becomes narrower at observed times, where the model is directly constrained by the available observations, and wider in the intervals without observations.
The DAPS posterior, on the other hand, remains wide throughout the assimilation window. In fact, the samples generated by DAPS do not seem to be affected by the observations at all and resemble prior samples rather than posterior ones. These results thus show the advantage of using an ensemble-based prior for sampling~$\hat{\boldsymbol{z}}_{0|\Psi}$ over the Langevin dynamics employed in the original DAPS scheme, which assumes an isotropic Gaussian prior.
However, for this representative ground truth trajectory, the spread in the posterior ensemble is noticeably smaller for DAPS-e compared to both DPS and the particle filter, despite the use of the stochastic EnKF formulation employed in DAPS-e.

\begin{figure}[!t]
    \centering
    \includegraphics[width=\linewidth]{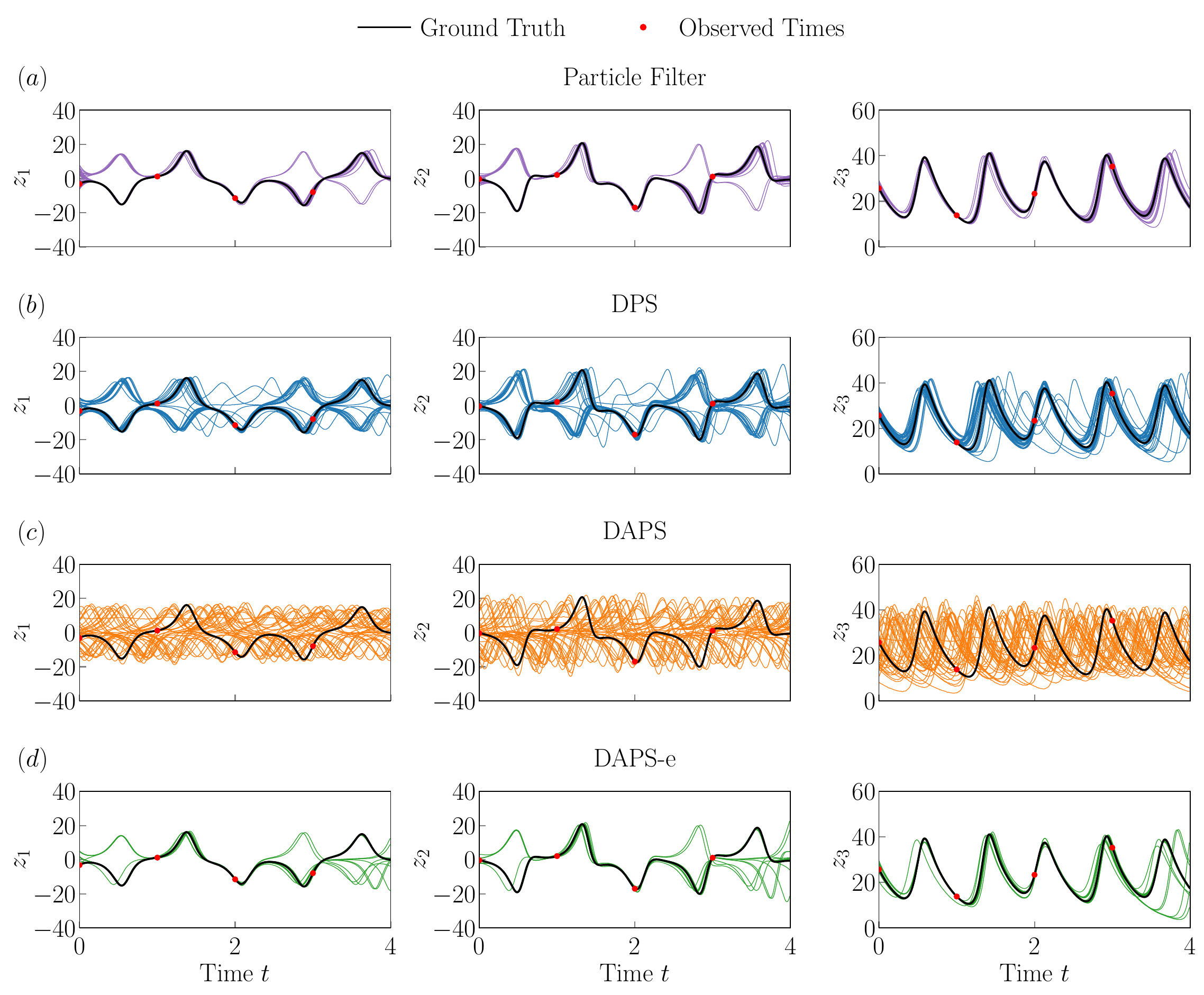}
    \caption{Representative posterior trajectories for the signed square operator~$\mathcal{H}(z)=z|z|$ generated by~$(a)$ the particle filter,~$(b)$ diffusion posterior sampling (DPS),~$(c)$ decoupled annealed posterior sampling (DAPS) and~$(d)$ the proposed method, DAPS with an ensemble-based prior covariance (DAPS-e).}
    \label{fig:posterior_samples_signed_square}
\end{figure}
\begin{figure}[!t]
    \centering
    \includegraphics[width=\linewidth]{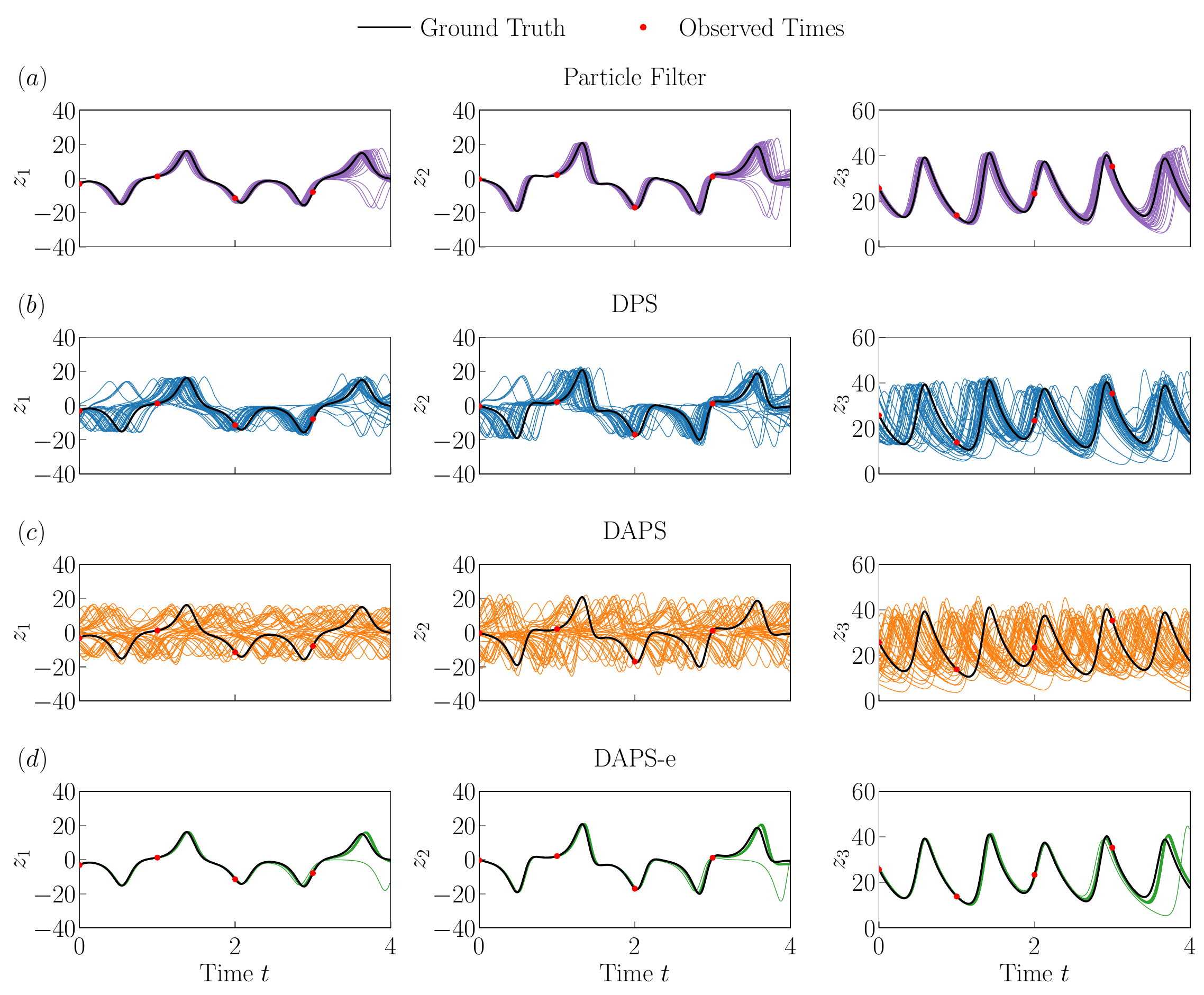}
    \caption{Representative posterior trajectories for the arctangent operator~$\mathcal{H}(z)=\operatorname{arctan}(z)$ generated by~$(a)$ the particle filter,~$(b)$ diffusion posterior sampling (DPS),~$(c)$ decoupled annealed posterior sampling (DAPS) and~$(d)$ the proposed method, DAPS with an ensemble-based prior covariance (DAPS-e).}
    \label{fig:posterior_samples_arctan}
\end{figure}
\begin{figure}[!t]
    \centering
    \includegraphics[width=0.85\linewidth]{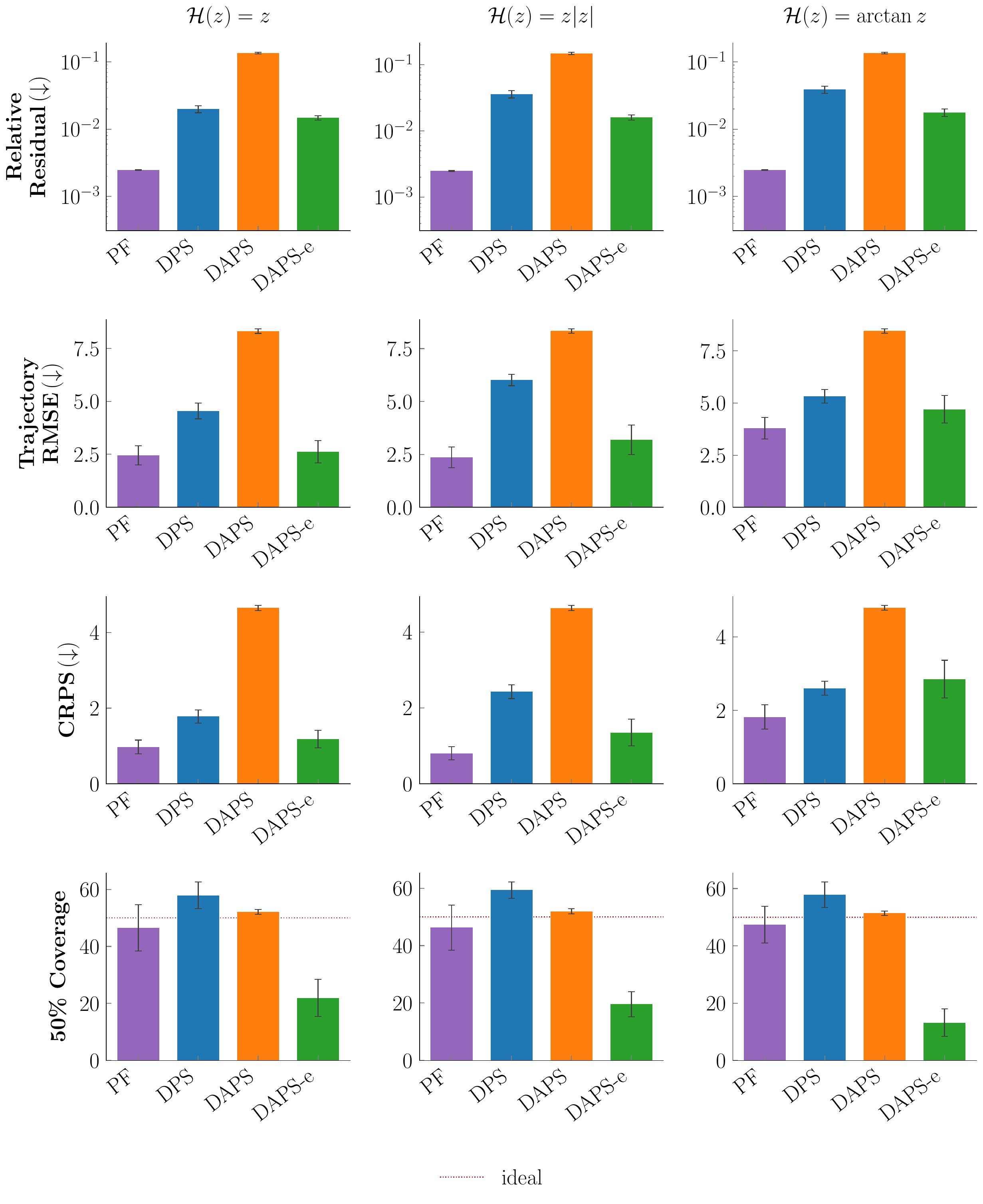}
     \caption{Summary of the quantitative evaluation of the posterior samples for the three operators considered: the identity operator (left), the signed square operator (middle), and the arctangent operator (right). We report the relative residual, the trajectory root mean squared error (RMSE), continuous ranked probability score (CRPS) and 50\% coverage.
     The arrow~$\downarrow$ is used to indicate that the lower the better for the respective metric.
     The metrics are averaged over 50 distinct trajectories. Error bars indicate the 95\% confidence interval of the means.}
    \label{fig:metrics_summary}
\end{figure}

DAPS-e shows similarly good qualitative agreement for the more challenging nonlinear observation operators. DAPS again shows little alignment with the ground truth for either the signed square or the arctangent operator, whereas the DAPS-e trajectories follow the ground truth closely (Figures~\mbox{\ref{fig:posterior_samples_signed_square}} and~\mbox{\ref{fig:posterior_samples_arctan}}). However, the DAPS-e ensemble is noticeably narrower than in the linear case, particularly for the signed square operator. For the arctangent operator, DAPS-e again shows limited spread around the ground truth, while DPS exhibits substantially greater uncertainty. This is consistent with the weakening of DPS guidance as the arctangent operator saturates (Section~\mbox{\ref{subsec:setup_lorenz}}). The particle filter also exhibits substantial weight degeneracy, particularly for the nonlinear operators, but its large particle populations yield effective sample sizes of order~$10^4$--$10^5$ on average with comparatively limited path degeneracy (Appendix~\mbox{\ref{sec:A3_pf}}), making it a useful qualitative baseline for assessing posterior spread.

\begin{figure}[!t]
    \centering    \includegraphics[width=0.9\linewidth]{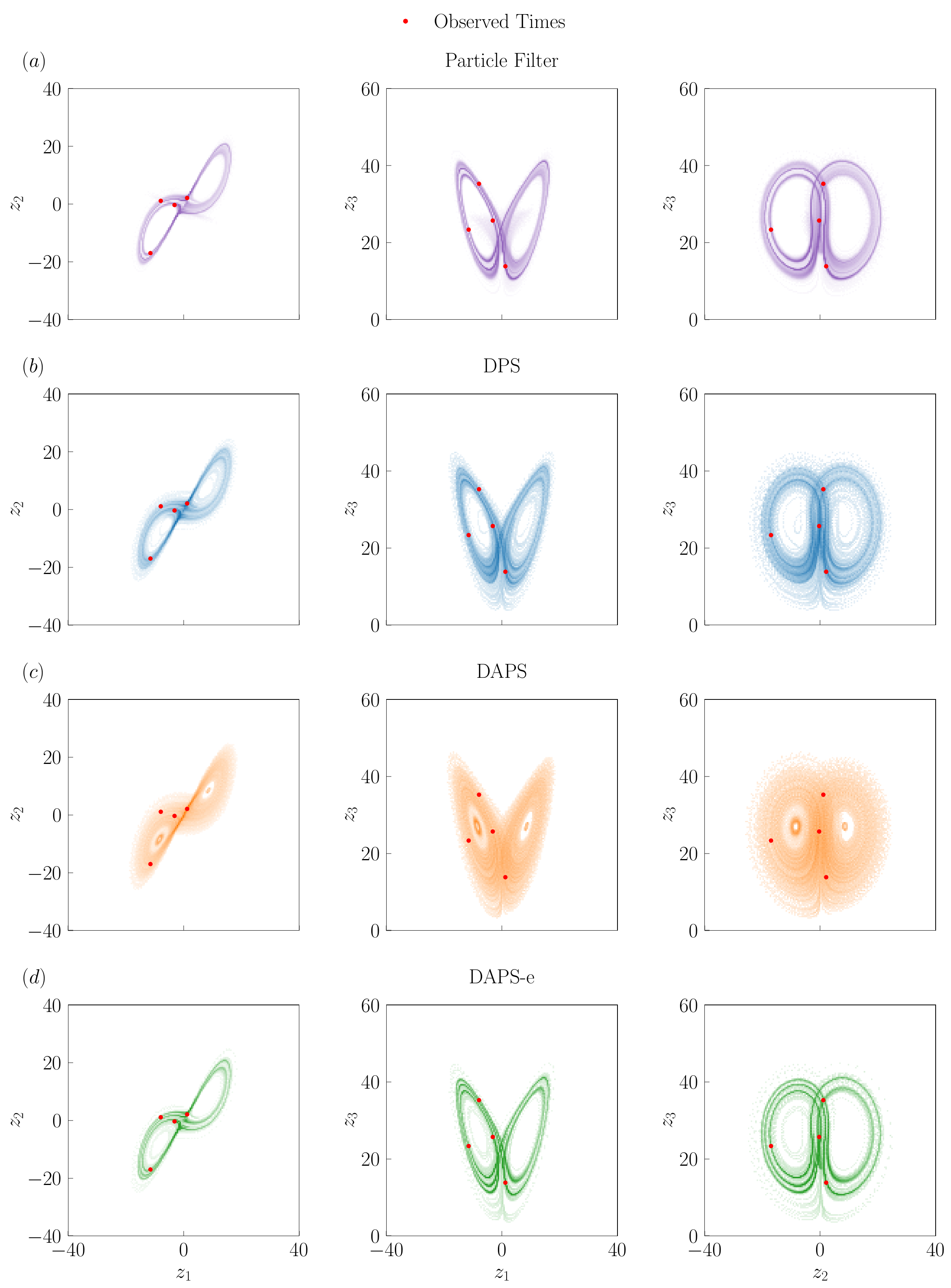}
    \caption{Pairwise joint distributions of the Lorenz-63 state variables for the signed
    square operator~$\mathcal{H}(z) = z\lvert z \rvert$. Rows show the posterior samples
    generated by~$(a)$ the particle filter,~$(b)$ diffusion posterior sampling (DPS),
   ~$(c)$ decoupled annealed posterior sampling (DAPS), and~$(d)$ the proposed method, DAPS with an
    ensemble-based prior covariance (DAPS-e). Columns show the
   ~$(z_1,z_2)$,~$(z_1,z_3)$ and~$(z_2,z_3)$ projections. Red dots mark the true state at
    the observed times.}
    \label{fig:joint}
\end{figure}

The quantitative evaluation shows that DAPS-e produces accurate and dynamically
consistent posterior samples for both linear and nonlinear observation operators,
but underestimates their uncertainty. We summarize the assessment in
Figure~\ref{fig:metrics_summary} for each operator, reporting the relative equation
residual, the trajectory root mean squared error (RMSE), the continuous ranked
probability score (CRPS) and the 50\% empirical coverage. The relative equation
residual measures the consistency of the generated trajectories with the Lorenz-63
dynamics, and the trajectory RMSE quantifies the error of the ensemble mean with
respect to the true trajectory. The CRPS assesses the ensemble as a whole, rewarding
both accuracy and an appropriate spread, while the empirical coverage measures how
often the true state lies within the central 50\% of the ensemble and thus assesses
whether the ensemble spread is appropriate. Precise definitions of all metrics are
given in Appendix~\ref{sec:A3}. The metrics are averaged over 50 test trajectories,
with 95\% confidence intervals of the means indicated in the figure.
DAPS-e outperforms the original DAPS in the relative residual, trajectory RMSE and
CRPS for all three observation operators. Compared with DPS, DAPS-e yields a lower
equation residual for all operators, with the difference becoming more pronounced for
the nonlinear operators, indicating that DAPS-e better preserves the governing
dynamics while assimilating the observations. For the identity and signed square
operators, DAPS-e also yields substantially lower RMSE and CRPS than DPS, approaching
the values of the particle filter. For the arctangent operator, DAPS-e yields a lower
residual and RMSE than DPS, while its CRPS is somewhat higher.
The empirical coverage indicates that the ensembles generated by DAPS-e are too narrow, confirming the conclusion drawn from the posterior trajectories in Figures~\ref{fig:posterior_samples_linear}--\ref{fig:posterior_samples_arctan}.
For all three operators, the true state lies within the central 50\% of the ensemble
for only 13--22\% of the time, well below the nominal value. The coverage of DPS and
the particle filter, in contrast, is close to the nominal value. The coverage of DAPS
is also close to 50\%, but this does not indicate a good posterior: since DAPS
effectively samples from the prior, and the true trajectory is itself a sample from
this prior, nominal coverage is expected regardless of accuracy. Overall, DAPS-e
generally improves the accuracy and dynamical consistency of the posterior samples
relative to DPS, whereas its ensemble spread underestimates the posterior
uncertainty. We reiterate that the guidance strength in DPS was tuned specifically
for this problem to a value one order of magnitude lower than that typically used in
image inverse problems, whereas the same value of~$c$ was used for DAPS-e in both
test cases.

The observed under-dispersion can be primarily attributed to the repeated use of the EnKF analysis within DAPS-e.
At each annealing level, DAPS requires samples from~$p(\boldsymbol{z}_0\mid\boldsymbol{z}_\tau,\boldsymbol{\Psi})$, whereas DAPS-e replaces this conditional sampling step with a stochastic EnKF update. Although the stochastic EnKF reproduces the Kalman analysis covariance in the linear--Gaussian setting, it is not an exact posterior sampler in general. A related contraction has been reported for ensemble Kalman guidance (EnKG), where the Kalman correction acts as an optimization step and drives the ensemble toward consensus~\mbox{\citep{zheng2025ensemble}}. In DAPS-e, the approximate EnKF correction is applied repeatedly along the annealing path, so its contraction can accumulate across successive levels. Re-noising restores part of the ensemble spread between updates, but does not guarantee that the complete denoise--assimilate--re-noise transition preserves the target posterior. The accumulated contraction therefore provides a natural explanation for the maximum a-posteriori (MAP)-like, under-dispersed ensembles observed here. Finite-ensemble sampling error, the resulting rank-deficient covariance, and the step-size limiter can further contribute to the loss of spread.

DAPS-e also preserves the geometry of the attractor in the joint distributions of the state variables, although, as for the marginals, its spread is too narrow. Figure~\ref{fig:joint} shows the pairwise
joint distributions of $(z_1,z_2)$,~$(z_1,z_3)$ and~$(z_2,z_3)$ for the signed square
operator. The two-dimensional projections of the attractor are recovered by DAPS-e as a
set of thin, well-separated filaments,
indicating that the posterior is concentrated on a small number of trajectories
consistent with both the attractor geometry and the observations. DAPS likewise preserves the coarse
geometry of the attractor, the two lobes and their low-density centers remaining
identifiable, but spreads its mass across and beyond them, so that its joint
distributions resemble the prior rather than a
posterior constrained by the observations. The limited posterior spread of DAPS-e is once again visible in the joint distribution, as the filaments are noticeably narrower than those of DPS.

\subsection{Rayleigh--Bénard Convection}
\begin{figure}[!t]
    \centering    \includegraphics[width=\linewidth]{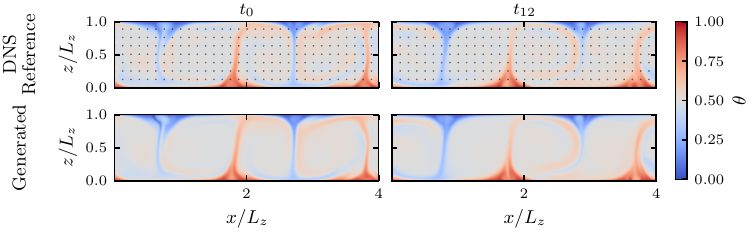}
    \caption{Temperature contours of two snapshots from the same trajectory separated by~$3t_f$. We show the DNS fields on the top row and the field generated by DAPS-e on the bottom. The observed locations are indicated with dots.}
    \label{fig:contours}
\end{figure}

Since turbulence is chaotic, a posterior sample that differs pointwise from the DNS realization can be an equally plausible realization of the same flow. The samples are therefore assessed by the statistics they reproduce rather than by pointwise error~\mbox{\citep{steinbrenner2026turbulence}}; pointwise agreement is meaningful only at the observation locations, where it measures consistency with the data. The snapshots of the generated temperature fields in Figure~\ref{fig:contours} accordingly serve only as a visual check that the guidance steers the samples toward the observed realization. Beyond containing the thermal plumes and convection rolls typical of Rayleigh--B\'{e}nard convection, the generated fields resemble the particular DNS realization that was observed. The physical consistency of the generated fields is assessed through the statistical quantities that follow.

\begin{table}[!b]
    \caption{Wall Nusselt numbers for the DNS and the posterior samples generated by DAPS-e. The relative error from DNS is reported for each wall.
    The top--bottom asymmetry quantifies departure from symmetric wall heat transfer and is defined as $|\mathrm{Nu}_{\mathrm{top}}-\mathrm{Nu}_{\mathrm{bottom}}|/[(\mathrm{Nu}_{\mathrm{top}}+\mathrm{Nu}_{\mathrm{bottom}})/2]$.}
    \label{tab:nusselt}
    \begin{tabular}{lccc}
        \toprule
        & DNS & DAPS-e & Relative error from DNS \\
        \midrule
        Top-wall $\mathrm{Nu}_{\mathrm{top}}$    & 17.31 & 16.32 & $5.7$\% \\
        Bottom-wall $\mathrm{Nu}_{\mathrm{bottom}}$ & 17.26 & 16.70 & $3.2$\% \\
        Top--bottom asymmetry      & 0.3\% & 2.3\% & -- \\
        \bottomrule
    \end{tabular}
\end{table}

The posterior samples generated by DAPS-e show good overall agreement with the DNS in terms of Nusselt number as well as mean and root mean squared (RMS) profiles.
We define the global Nusselt number at the walls as
\begin{equation}
        \left. \mathrm{Nu}_{\mathrm{wall}} = - \frac{L_z}{\Delta\theta}\, \overline{\frac{\partial \theta}{\partial z}}\right|_{\text{wall}},
\end{equation}
where~$\Delta \theta=1$ is the temperature difference between the two plates. The Nusselt numbers evaluated at the top and bottom walls are reported in Table~\ref{tab:nusselt} for the DAPS-e posterior samples and the DNS. The Nusselt numbers of the posterior samples lie 5.7\% and 3.2\% below the DNS values at the top and bottom walls, respectively, showing that the global heat-transfer rate is captured reasonably well. Owing to the symmetry of the Rayleigh--Bénard setup, the top- and bottom-wall Nusselt numbers should agree in the statistically converged limit. The posterior samples show a small top--bottom asymmetry of 2.3\%, compared with 0.3\% for the DNS, indicating a modest departure from the expected statistical symmetry.
\begin{figure}[!t]
    \centering
    \includegraphics[width=\linewidth]{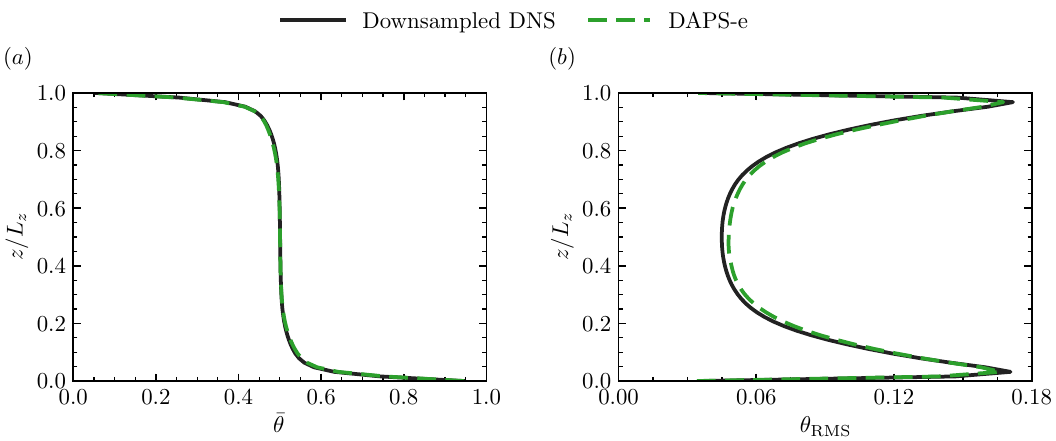}
    \caption{The mean~$(a)$ and root mean squared (RMS)~$(b)$ temperature profile of the DNS fields and the posterior samples generated by DAPS-e.}
    \label{fig:mean-RMS}
\end{figure}
\begin{figure}[!t]
    \centering
    \includegraphics[width=\linewidth]{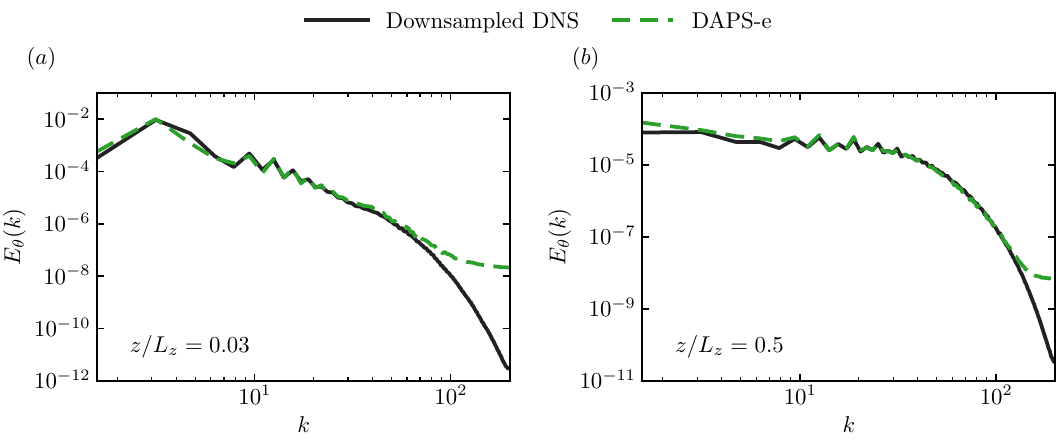}
    \caption{Temperature spectra at~$(a)$ near the bottom wall, i.e., $z/L_z=0.03$, and~$(b)$ at mid-height, i.e.,~$z/L_z=0.5$, for the DNS fields and the posterior samples generated by DAPS-e.}
    \label{fig:spectra}
\end{figure}
The mean temperature profile of the DAPS-e samples agrees closely with the DNS, and the RMS profile departs from it only slightly in the bulk (Figure~\ref{fig:mean-RMS}). The profiles exhibit the characteristic structure of
Rayleigh--B\'enard convection: the mean temperature profile shows steep gradients
within the thermal boundary layers near the walls, while the RMS profile displays
two near-wall peaks and an approximately uniform plateau in the bulk.
\begin{figure}[!t]
    \centering
    \includegraphics[width=0.5\linewidth]{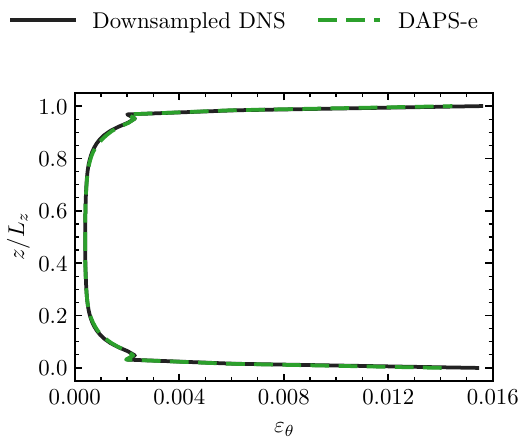}
    \caption{Thermal dissipation profile of the DNS fields and posterior samples generated by DAPS-e.}
    \label{fig:dissipation_profile}
\end{figure}

The samples generated by DAPS-e reproduce the large-scale temperature
fluctuations reasonably well, but discrepancies remain at smaller scales. To
quantify how thermal fluctuations are distributed across scales, we examine the
temperature spectra in Figure~\ref{fig:spectra}, evaluated both near the wall
and at mid-height.
It can be seen that the posterior samples match the DNS spectra in the lower wavenumbers,
but exhibit excess energy at high wavenumbers, corresponding to the small-scale range.
\citet{sardar2024spectrally} reported a similar high-wavenumber energy buildup in flow fields generated by diffusion models.
The effect of these
high-wavenumber errors is amplified in gradient-based quantities such as the
thermal dissipation rate,~$\varepsilon_\theta(z)=\kappa\overline{|\nabla
\theta'|^2}$, where~$\kappa=1/\sqrt{\mathrm{Ra} \cdot \mathrm{Pr}}$ is the nondimensional thermal
diffusivity. The vertical profile of~$\varepsilon_\theta$, shown in
Figure~\ref{fig:dissipation_profile}, confirms this sensitivity. While the dissipation rate of the posterior samples exhibits very good agreement with the DNS in the bulk flow,
posterior samples underpredict the thermal dissipation near the walls, where the
sharpest temperature gradients occur. The spectra and dissipation profiles thus
provide complementary evidence that DAPS-e recovers the large-scale thermal
structure but still struggles to reproduce the near-wall, small-scale gradient
content of the flow.
\section{Conclusion}\label{sec:conclusion}
This work set out to ask how far a diffusion-based assimilation scheme can retain the Bayesian structure of data assimilation, so that the weight given to observations follows from their prescribed error covariance rather than from a tuned hyperparameter. To that end, we combined decoupled annealed posterior sampling (DAPS) with an ensemble estimate of the prior covariance, yielding DAPS with an ensemble-based prior covariance (DAPS-e). DAPS moves the assimilation step to the clean state, where the observation likelihood is naturally defined and the prescribed observation-error covariance can be incorporated explicitly. The incorporation of the EnKF into DAPS replaces the isotropic Gaussian covariance approximation used in the original method with an ensemble-based estimate of the uncertainty structure of the clean-state samples. DAPS-e therefore moves in the direction set out at the start of this work, in two respects: the likelihood is evaluated in the clean-state space where the observation model applies, and the prior covariance is estimated from the ensemble rather than prescribed. This is one route among others, and not necessarily the best one. However, the objective itself seems to us worth pursuing. For the Lorenz-63 system, with both linear and nonlinear observation operators, DAPS-e yields lower equation residuals and trajectory errors than DPS and the original DAPS, although its ensemble is too narrow. For two-dimensional Rayleigh–Bénard convection, the framework generates physically plausible posterior samples from sparse observations, matching the DNS in several low-order statistics, although discrepancies remain at small scales. These results indicate the potential of the framework for more challenging turbulent flows.

While the present results demonstrate the potential of the proposed ensemble-based DAPS framework, several issues remain before the framework can be applied in practical settings. First, DAPS-e underestimates posterior uncertainty: in the Lorenz-63 case, the nominal 50\% ensemble interval achieves only 13--22\% empirical coverage. This under-dispersion reflects the repeated use of EnKF updates for conditional sampling; their contraction can accumulate across annealing levels and need not preserve the posterior spread. Improving this conditional sampling step is therefore an important direction for future work. Second, both test cases use synthetic observations extracted from numerical simulations; future work should therefore assess the framework using measurements from physical systems and in the presence of model error. Finally, the present study applies DAPS-e entirely in physical space. Since generative models for high-dimensional turbulent flows are often formulated in a lower-dimensional latent space, extending the framework to latent space diffusion models will be important for improving its scalability to more realistic flow configurations.

\vspace{\baselineskip}
\noindent\textbf{Author Contributions}
All authors designed the research. B.T. and Z.L. developed the methodology and conducted the simulations, model training, and analysis. All authors contributed to the manuscript preparation. H.X. supervised the research. All authors reviewed and approved the manuscript.

\vspace{\baselineskip}
\noindent\textbf{Funding}
The authors acknowledge funding from the Deutsche Forschungsgemeinschaft (DFG, German Research Foundation) under Germany's Excellence Strategy -- EXC~2075 -- 390740016, and support from the Stuttgart Center for Simulation Science (SimTech). B.T. acknowledges funding from the DFG under Grant No.~551388164. B.T. is also supported by the Carl Zeiss Foundation (CZS Project Number P2021-04012).

\vspace{\baselineskip}
\noindent\textbf{Data Availability}
The code and the data supporting the findings of this study will be made publicly available at~\mbox{\url{https://github.com/ITLR-DDSim/daps-e-data-assimilation}} upon acceptance.

\section*{Declarations}
\noindent\textbf{Competing Interests}
The authors declare no competing interests.

\begin{appendices}
\section{Neural Network Architectures and
Hyperparameters}\label{sec:A1}
\subsection{Training Objective}\label{sec:A1_training}
The score network~$\boldsymbol{s}_\phi$ of Section~\ref{subsec:diffusion_models} is trained by minimizing the denoising score matching loss
\begin{equation}
    \mathcal{L}_{\mathrm{DSM}}(\phi)
    =
    \mathbb{E}_{\tau,\boldsymbol{z}_0,\boldsymbol{z}_\tau}
    \left[
        \lambda(\tau)
        \left\|
            \boldsymbol{s}_\phi(\boldsymbol{z}_\tau,\tau)
            -
            \nabla_{\boldsymbol{z}_\tau}
            \log p
            (\boldsymbol{z}_\tau\mid\boldsymbol{z}_0)
        \right\|_2^2
    \right],
\end{equation}
    where~$\lambda(\tau)$ is a pseudo-time-dependent weighting factor. With enough training data and model capacity, the minimizer of this loss converges to the marginal score~$\nabla_{\boldsymbol{z}_\tau}\log p(\boldsymbol{z}_\tau)$~\citep{song2021scorebased}.

In this work, we adopt the learning objective of the denoising diffusion probabilistic model (DDPM;~\citealp{ho2020denoising}), which parameterizes the problem in terms of the added noise instead of the score function. Using Equation~\ref{eq:forward_kernel}, we can sample the forward kernel as
\begin{equation}
   \boldsymbol{z}_\tau=\mu_\tau\boldsymbol{z}_0+\sigma_\tau\boldsymbol{\epsilon}_\tau, \, \boldsymbol{\epsilon}_\tau\sim \mathcal{N}(0, \boldsymbol{I}).
\end{equation}
Since~$p(\boldsymbol{z}_\tau\mid\boldsymbol{z}_0)$ is Gaussian, its score is given by~$\nabla_{\boldsymbol{z}_\tau}
\log p
(\boldsymbol{z}_\tau\mid\boldsymbol{z}_0)=-\frac{\boldsymbol{\epsilon}_\tau}{\sigma_\tau}$. The network can hence be trained to predict the added noise~$\boldsymbol{\epsilon}_\tau$ instead of the score function in this formulation. The final form of the loss function then reads
\begin{equation}
    \mathcal{L}_{\mathrm{simple}}(\phi)
    =
    \mathbb{E}_{\tau,\boldsymbol{z}_0,\boldsymbol{\epsilon}_{\tau}}
    \left[
        \left\|
            \boldsymbol{\epsilon}_\tau
            -
            \boldsymbol{\epsilon}_\phi
            \left(\boldsymbol{z}_\tau,
                \tau
            \right)
        \right\|_2^2
    \right],
\end{equation}
where we drop the weighting factor~$\lambda(\tau)$, as in the simplified DDPM training objective~\citep{ho2020denoising}.
\setcounter{equation}{0}
\subsection{Lorenz-63 System}
We parameterize the diffusion noise predictor~$\boldsymbol{\epsilon}_{\phi}(\boldsymbol{\boldsymbol{z}}_\tau,\tau)$ using a one-dimensional convolutional UNet operating along the physical time dimension of the Lorenz-63 trajectories. The three state variables are treated as input channels, while all convolutions and down-/upsampling operations act along the trajectory dimension. The network consists of three resolution levels with skip connections between the encoder and decoder. The pseudo-time~$\tau$ is represented by a sinusoidal embedding followed by a multilayer perceptron and is injected into each residual block through feature-wise linear modulation. We use the variance-preserving SDE with a linear noise schedule. Since the architecture is fully convolutional in physical time, its parameters are independent of the trajectory length~$L$, allowing the same network to process trajectories of different lengths. Additional details regarding the network architecture are provided in Table \ref{tab:lorenz_unet}. We train the model using the AdamW optimizer~\citep{loshchilov2018decoupled} with a learning rate of~$3\times10^{-4}$ and a weight decay of~$1\times10^{-4}$. During training, we maintain an exponential moving average of the model parameters with a decay rate of~$0.999$, which is used to improve training stability and generalization.
For the diffusion model, we follow the DDPM framework of \citet{ho2020denoising} and use their hyperparameters for the diffusion model. Specifically, we use 1000 diffusion steps and a linear noise schedule.
\begin{table}[!h]
\centering
\caption{Architecture of the one-dimensional UNet used as the diffusion noise predictor for Lorenz-63 trajectories.}
\label{tab:lorenz_unet}
\begin{tabular}{ll}
\hline
\textbf{Component} & \textbf{Configuration} \\
\hline
Input & $3 \times L$ Lorenz trajectory \\
Resolution levels & $3$ \\
Channel dimensions & $64,\ 128,\ 256$ \\
Residual blocks per level & $2$ \\
Temporal downsampling factor per level & $2$ \\
Pseudo-time embedding & Sinusoidal embedding, dimension~$128$ \\
Activation & SiLU \\
Normalization & Group normalization \\
Noise schedule & Variance-preserving SDE, linear schedule \\
Output & $3 \times L$ predicted noise \\
\hline
\end{tabular}
\end{table}
\subsection{Rayleigh--Bénard Convection}
For Rayleigh--B\'enard convection, we follow the Markov-blanket approach of the
framework of \citet{rozet2023score}. Given a trajectory
$\boldsymbol{z}^{1:L}=\{\boldsymbol{z}^1,\ldots,\boldsymbol{z}^L\}$ and its
noised counterpart~$\boldsymbol{z}^{1:L}_\tau$, the score of the noised
trajectory is
$\boldsymbol{s}(\boldsymbol{z}^{1:L}_\tau)
=\nabla_{\boldsymbol{z}^{1:L}_\tau}\log p(\boldsymbol{z}^{1:L}_\tau)$.
Rather than evaluating this score globally over the full trajectory, the
Markov-blanket approximation assumes that the score contribution at time index~$i$ depends only on a local temporal neighborhood
$\{i-k_{M},\ldots,i+k_{M}\}$, where $k_{M}$ is the temporal half-window size. This allows scores for arbitrarily long trajectories to be
assembled from overlapping windows of length~$W=2k_{M}+1$. In this work, we use~$k_{M}=3$.
\begin{table}[!t]
\centering
\caption{Architecture of the two-dimensional UNet used as the diffusion noise
predictor for Rayleigh--B\'enard convection.}
\label{tab:rbc_unet}
\begin{tabular}{ll}
\hline
\textbf{Component} & \textbf{Configuration} \\
\hline
Input & $7$ temperature frames, $256 \times 64$ ($x \times z$) grid \\
Auxiliary channel & Fixed normalized-height map \\
Resolution levels & $3$ \\
Channels & $64,\ 128,\ 256$ \\
Residual blocks & $3$ per level \\
Convolutions & $3 \times 3$ \\
Down/up-sampling & Factor~$2$ per level; nearest-neighbor upsampling \\
Skip connections & Between matching encoder--decoder levels \\
Pseudo-time embedding & Sinusoidal, dimension~$64$ \\
Time conditioning & Additive learned projection in each residual block \\
Activation / normalization & SiLU / layer normalization \\
Noise schedule & VP-SDE with cosine schedule \\
Output & Predicted noise over the~$7$-frame window \\
\hline
\end{tabular}
\end{table}

The local score model is implemented as a two-dimensional UNet noise predictor
acting on the standardized dimensionless temperature field. The temporal window
is folded into the channel dimension, so that the~$W$ single-channel temperature
frames are provided as input channels. Since Rayleigh--B\'enard convection is
not translation invariant in the vertical direction, a fixed normalized-height
channel is concatenated to the input to distinguish the hot lower plate from the
cold upper plate. The network is trained as a noise estimator for a
variance-preserving SDE with a cosine noise schedule and predicts the noise over
the full input window. It is trained with the AdamW optimizer
with a learning rate of $2\times10^{-4}$, an
exponential learning-rate schedule and a weight decay of $1\times10^{-3}$. The architectural details are summarized in
Table~\ref{tab:rbc_unet}.

\section{Computation of the Kalman Gain Matrix}\label{sec:A2}
Given an ensemble of clean-state estimates
$\hat{\boldsymbol{Z}}_0 =
[\hat{\boldsymbol{z}}_0^{(1)},\ldots,\hat{\boldsymbol{z}}_0^{(N_\mathrm{ens})}]$,
we define the normalized state anomaly matrix as
\begin{equation}
    \boldsymbol{S}_z
    =
    \frac{1}{\sqrt{N_\mathrm{ens}-1}}
    \left[
    \hat{\boldsymbol{z}}_0^{(1)}-\bar{\boldsymbol{z}}_0,
    \ldots,
    \hat{\boldsymbol{z}}_0^{(N_\mathrm{ens})}-\bar{\boldsymbol{z}}_0
    \right],
\end{equation}
where~$\bar{\boldsymbol{z}}_0$ is the ensemble mean. Similarly, we evaluate
the observation operator for every ensemble member and define the normalized
observation anomaly matrix
\begin{equation}
    \boldsymbol{S}_h
    =
    \frac{1}{\sqrt{N_\mathrm{ens}-1}}
    \left[
    \mathcal{H}(\hat{\boldsymbol{z}}_0^{(1)})-\bar{\boldsymbol{h}},
    \ldots,
    \mathcal{H}(\hat{\boldsymbol{z}}_0^{(N_\mathrm{ens})})-\bar{\boldsymbol{h}}
    \right],
\end{equation}
where
\begin{equation}
    \bar{\boldsymbol{h}}
    =
    \frac{1}{N_\mathrm{ens}}
    \sum_{i=1}^{N_\mathrm{ens}}
    \mathcal{H}(\hat{\boldsymbol{z}}_0^{(i)}).
\end{equation}
The sample state-observation and observation covariances are therefore
$\boldsymbol{P}_{zh}=\boldsymbol{S}_z\boldsymbol{S}_h^\top$ and
$\boldsymbol{P}_{hh}=\boldsymbol{S}_h\boldsymbol{S}_h^\top$,
respectively, giving the Kalman gain
\begin{equation}
    \boldsymbol{K}
    =
    \boldsymbol{S}_z\boldsymbol{S}_h^\top
    \left(
    \boldsymbol{S}_h\boldsymbol{S}_h^\top
    +
    \boldsymbol{R}
    \right)^{-1}.
\end{equation}
No covariance localization or covariance inflation is applied in the present implementation. Consequently, the covariance represented by an ensemble of~$N_\mathrm{ens}$ members has rank at most~$N_\mathrm{ens}-1$.
\setcounter{equation}{0}
\section{Evaluation Metrics for the Lorenz-63 System}\label{sec:A3}
\subsection{Relative Residual}
We define the relative residual of the Lorenz-63 equations as
\begin{equation}
    r=\frac{\lVert \dot{\boldsymbol{z}} - \boldsymbol{f}_{\mathrm{L}}(\boldsymbol{z})\rVert}{\lVert \boldsymbol{f}_{\mathrm{L}}(\boldsymbol{z})\rVert},
\end{equation}
where~$\boldsymbol{z}=[z_1, z_2, z_3]^\top$ is the Lorenz-63 state and~$\boldsymbol{f}_{\mathrm{L}}(\boldsymbol{z})$ is the right-hand side of the Lorenz-63 equations (Equations~ \ref{eq:lorenz1}--\ref{eq:lorenz3}).
\subsection{Trajectory RMSE}
The trajectory root mean squared error (RMSE) is given by
\begin{equation}
    \mathrm{RMSE}
=
\sqrt{
\frac{1}{Ld}
\sum_{t=1}^{L}
\sum_{j=1}^{d}
\left(
\bar{z}_{t,j}-z^\ast_{t,j}
\right)^2
},
\end{equation}
where~$L$ is the trajectory length,~$d$ is the number of state variables,~$\bar{\boldsymbol{z}}$ is the ensemble mean and~$\boldsymbol{z}^*$ is the ground truth trajectory. For the Lorenz-63 system, we have~$L=400$ and~$d=3$.
\subsection{Continuous Ranked Probability Score}
The continuous ranked probability score is widely used to quantify the accuracy of probabilistic forecasters~\citep{gneiting2007strictly}. It is defined as
\begin{equation}
    \mathrm{CRPS} = \frac{1}{N_\mathrm{ens}}\sum_{i=1}^{N_\mathrm{ens}}
    \left|z^{(i)} - z^{*}\right|
    - \frac{1}{2N_\mathrm{ens}^{2}}\sum_{i=1}^{N_\mathrm{ens}}
      \sum_{j=1}^{N_\mathrm{ens}}
      \left|z^{(i)} - z^{(j)}\right|.
\end{equation}
CRPS is a strictly proper scoring rule, meaning that it is minimized in expectation by the true distribution of the target variable~$\boldsymbol{z}^*$. We compute the CRPS separately for each state variable and report the average over all three state variables in this work.

\subsection{Empirical Coverage}
The empirical coverage assesses whether the spread of the ensemble is consistent with
its actual error. For an ensemble of $N_\mathrm{ens}$ members $\boldsymbol{z}^{(i)}$, let
$q^{p}_{t,j}$ denote the empirical $p$-quantile of the ensemble values
$\{z^{(i)}_{t,j}\}_{i=1}^{N_\mathrm{ens}}$ of state variable $j$ at time $t$. The
central interval of nominal level $\gamma$ is bounded by the quantiles
$p_{\mathrm{lo}} = (1-\gamma)/2$ and $p_{\mathrm{hi}} = (1+\gamma)/2$, and the
empirical coverage is the fraction of time instants and state variables at which the
ground truth lies within this interval,
\begin{equation}
    \mathrm{Cov}_{\gamma} = \frac{1}{Ld}\sum_{t=1}^{L}\sum_{j=1}^{d}
    \mathbb{I}\!\left[\, q^{p_{\mathrm{lo}}}_{t,j} \le z^{*}_{t,j}
    \le q^{p_{\mathrm{hi}}}_{t,j} \,\right],
\end{equation}
where $\mathbb{I}[\cdot]$
equals one if the condition holds and zero otherwise. In this work we use
$\gamma = 0.5$, i.e., the interval between the 25th and 75th percentiles of the
ensemble. If the ensemble is a sample of the correct posterior distribution, the
ground truth is itself a draw from this distribution at every time instant, and the
expected coverage equals the nominal level $\gamma$. Values below $\gamma$ indicate
that the ensemble is too narrow, i.e., it underestimates the uncertainty, and values
above $\gamma$ indicate that it is too wide. The coverage is computed for each test
trajectory and then averaged over all trajectories. Nominal coverage is a necessary
but not a sufficient condition for a good posterior: a sampler that ignores the
observations and draws from the prior also attains the nominal coverage, since the
true trajectory is a sample from the same prior.
\subsection{Particle-Filter Diagnostics}\label{sec:A3_pf}
We assess the quality of the particle
approximation using two diagnostics: the minimum normalized effective sample
size,
$\mathrm{ESS}/N_p = (N_p\sum_{i=1}^{N_p} w_i^2)^{-1}$,
where~$N_p$ is the number of particles and~$w_i$ are the normalized particle
weights, and the number of unique ancestors at the first assimilation cycle,
$n_a$, obtained by tracing a random sample of 2000 particles retained at the
final assimilation time backward through the resampling history. The normalized
ESS quantifies the effective fraction of particles contributing to the weighted
approximation, whereas~$n_a$ provides a measure of path degeneracy, or the loss of particle diversity through resampling.
 Across the 50 test trajectories, the mean minimum normalized ESS was~$10.6\%$,
$1.72\%$, and~$9.94\%$ for the identity, signed square, and arctangent operators,
respectively, while the mean number of unique initial ancestors among 2000
retained trajectories was 1687, 1429, and 1689. Considerable
trajectory-to-trajectory variability was observed, particularly for the
signed square operator, for which the normalized ESS ranged from~$0.027\%$ to
$14.5\%$ and the number of unique ancestors from 30 to 1991.
These results indicate substantial weight degeneracy, particularly for the
nonlinear observation operators, but comparatively limited loss of particle
diversity on average. For the signed square operator, 1429 of the 2000 retained
trajectories still originate from distinct particles at the first assimilation
cycle on average, although a few realizations show substantially stronger
collapse. Despite the low normalized ESS values, the large
particle populations still yield effective sample sizes of order~$10^4$--$10^5$ on
average.
\setcounter{equation}{0}

\section{Step-Size Limiter for the Ensemble Update}\label{sec:limiter}
The bound applied to the ensemble update in Section~\ref{subsec:daps_e} is implemented as follows.
The limited update is
\begin{equation}
    \Delta\hat{\boldsymbol{z}}_\tau
    =
    \eta_\tau
    \Delta\hat{\boldsymbol{z}}^\mathrm{raw}_\tau,
\end{equation}
where~$\Delta\hat{\boldsymbol{z}}^\mathrm{raw}_\tau
=
\hat{\boldsymbol{z}}_{0|\boldsymbol{\Psi}}
-
\hat{\boldsymbol{z}}_0$
denotes the raw EnKF update. The scaling coefficient is defined as
\begin{equation}
    \eta_\tau
    =
    \min\!\left(
    1,
    \frac{c\,\sigma_{\mathrm{eff},\tau}}
    {\left\lVert
    \Delta\hat{\boldsymbol{z}}_\tau^{\mathrm{raw}}
    \right\rVert_{\mathrm{RMS}}+\delta}
    \right),
\end{equation}
where~$c$ is a dimensionless hyperparameter, and~$\delta$ is a small constant introduced for numerical stability. The global RMS norm is calculated as
\begin{equation}
    \left\lVert
    \Delta\hat{\boldsymbol{z}}_\tau^{\mathrm{raw}}
    \right\rVert_{\mathrm{RMS}}
    =
    \sqrt{
    \frac{1}{N_\mathrm{ens} D}
    \sum_{i=1}^{N_\mathrm{ens}}
    \left\lVert
    \Delta\hat{\boldsymbol{z}}_{\tau}^{\mathrm{raw},(i)}
    \right\rVert_2^2
    },
\end{equation}
where~$N_\mathrm{ens}$ is the ensemble size and~$D$ is the number of entries of each generated trajectory. The limiter therefore constrains the RMS magnitude of the EnKF correction to be at most~$c\,\sigma_{\mathrm{eff},\tau}$.
When the raw update already satisfies this bound,~$\eta_\tau=1$ and the EnKF update is left unchanged. The update is therefore restricted to a ball of radius~$c\,\sigma_{\mathrm{eff}, \tau}$,
within which the linear--Gaussian description underlying the ensemble Kalman update is
taken to remain adequate. The idea is analogous to the trust-region methods used in
nonlinear least-squares optimization, which have also been applied to iterative
ensemble Kalman methods and ensemble Kalman inversion~\citep{chada2021}. It
differs from them in two respects. First, in a trust-region method such as the
Levenberg--Marquardt algorithm~\citep{more1978} the radius is adjusted at every iteration by comparing
the improvement that the local model predicted with the improvement actually obtained,
whereas here it is prescribed and set by the effective noise level~$\sigma_{\mathrm{eff}, \tau}$. Second, a trust-region method searches the ball for the step
that is best according to the local model, whereas we shorten the ensemble Kalman
update until it fits. The shortened update is the admissible step closest to the one
originally proposed and retains its direction, but it is not in general the best step
available within the ball.

\section{Computational Cost}\label{sec:A4}
For the Lorenz-63 case, all computations were performed on an NVIDIA RTX 5080 GPU. Training the diffusion model required approximately 15 minutes, while generating one posterior ensemble required approximately 9\,s with DAPS-e, compared with 19\,s for DPS. For the Rayleigh--B\'enard convection case, the diffusion model was trained on an NVIDIA RTX 6000 Ada Generation GPU for approximately 8 hours. Generating one DAPS-e posterior ensemble required approximately 80 minutes on the same hardware.

\end{appendices}

\bibliography{sn-bibliography}

\end{document}